\documentclass[a4paper,11pt]{article}
\pdfoutput=1 % if your are submitting a pdflatex (i.e. if you have
\usepackage{jheppub}
\usepackage[T1]{fontenc} % if needed
\usepackage[all]{xy}
\usepackage{rotating}
\usepackage{float}
\usepackage{tikz}
\usepackage{circuitikz}
\usepackage{tikz-network}
\usepackage{diagbox}
\usepackage{braket}
\usepackage{empheq}
\usepackage{tabularx}
\usepackage{multirow}
\usepackage{slashed}
\usepackage[normalem]{ulem}

\usepackage{tikz}
\usepackage{tikz-cd}
\usepackage{diagbox}
\usetikzlibrary{positioning}
\usetikzlibrary{calc}
\usetikzlibrary{decorations.pathreplacing,calligraphy}

\usepackage{xstring}
\usetikzlibrary{decorations.pathmorphing} 
\usetikzlibrary{decorations.markings} 
\usetikzlibrary{arrows} 
\usetikzlibrary{shapes} 
\usetikzlibrary{matrix} 
\usetikzlibrary{positioning} 
\usepackage[english]{babel} 
\usepackage[autostyle]{csquotes}
\usetikzlibrary{positioning,fit}
\usepackage[dvipsnames]{xcolor}
\tikzset{->-/.style={decoration={
  markings,
  mark=at position #1 with {\arrow{>}}},postaction={decorate}}}

   \let\d=\delta 
    
\let\l=\lambda \let\m=\mu \let\n=\nu   \let\r=\rho
\let\s=\sigma     
   \let\G=\Gamma \let\D=\Delta  \let\L=\Lambda
  
\def\CI{{\cal I}}

\def\CK{{\cal K}}

\def\CN{{\cal N}}
\def\CO{{\cal O}}
\def\CP{{\cal P}}

\def\CR{{\cal R}}
\def\CS{{\cal S}}
\def\CT{{\cal T}}

\def\Wg{\widetilde{g}}

\def\WK{\widetilde{K}}

\def\WO{\widetilde{O}}

\def\beq#1\eeq{\begin{align}#1\end{align}}

\makeatletter
\newcommand*{\rom}[1]{\expandafter\romannumeral #1}

\makeatother%For Roman Numeral

\title{3d $\mathcal{N}=$ 4 rank-0 SCFT from punctured lens space}

\abstract{{\it Gang-Kim-Stubbs} theory $\mathcal{T}_n$ --- a pioneering 3d bulk description of $M(2n+3,2)$ Virasoro minimal model as $\mathcal{N}=4$ rank-0 superconformal field theory upon topological A-twist --- is derived from the compactification of a pair of parallel M5-branes on {\it lens space} $L(2n+3,2)$ with a single vertex removed. From this perspective, we propose a family of 3d $\mathcal{N}=2$ abelian gauge theories arising from the punctured $L(2n+3,1)$ lens space whose infrared phases realize the unitary member in the Galois orbit of $M(2n+3,2)$ modular tensor category. We also conjecture self-mirror rank-0 fixed points from amphichirality condition of the lens space.}

\author[]{Sungjoon Kim}

\affiliation[]{Korea Institute for Advanced Study, 85 Hoegiro, Dongdaemun-Gu, Seoul 02455, Korea}

\emailAdd{sungjoon@kias.re.kr}

\begin{document} 
\preprint{KIAS-P26037}
\maketitle
\flushbottom

%%%%%%%%%%%%%%%%%%%%%%%%%%%%%%%%%%%%%%%%%%%%%%%%%%%%%%%%%%%%%%%%%%%%%%%%
%%%%%%%%%%%%%%%%%%%%%%%%%%%%%%%%%%%%%%%%%%%%%%%%%%%%%%%%%%%%%%%%%%%%%%%%

%%%%%%%%%%%%%%%%%%%%%%%%%%%%%%%%%%%%%%%%%%%%%%%%%%%%%%%%%%%%%%%%%%%%%%%%

\section{Introduction}
Following the discovery of {\it 3d rank-0} superconformal field theory (SCFT)~\cite{Gang:2018huc} --- which has neither Higgs nor Coulomb branch --- and an investigation of its relation to 3d non-unitary topological quantum field theories (TQFTs)~\cite{Gang:2021hrd}, this class of theories has attracted significant attention in a variety of contexts \cite{Gang:2023rei,Ferrari:2023fez,Dedushenko:2023cvd,Kim:2024dxu,Gaiotto:2024ioj,Go:2025ixu,ArabiArdehali:2024ysy,ArabiArdehali:2024vli,Creutzig:2024ljv,Kim:2025klh,Kim:2025rog,Nishinaka:2025ytu,Creutzig:2026ajk,Gang:2018wek,Garozzo:2019ejm,Cho:2020ljj,Assel:2022row,Gang:2022kpe,Choi:2022dju,Gang:2023ggt,Gang:2024tlp,Baek:2024tuo,Gang:2024loa,Jeong:2025xid,Baek:2025uev,Gang:2025ykf,Gang:2026iem,Yoshida:2026wvb,Jeong:2026dzz}. One striking feature of the 3d rank-0 SCFT is that it appears to have $\CN=4$ or $\CN=5$ supersymmetry without manifest lagrangian description. This amount of supersymmetry allows 3d topological A-twist~\cite{Kapustin:2010ag,Witten:1988ze} or B-twist~\cite{Blau:1996bx,Rozansky:1996bq}, capturing the non-unitary 3d TQFT structures. Equipped with the supersymmetric localization techniques \cite{Closset:2018ghr,Closset:2019hyt}, the rank-0 SCFT provides a convenient bulk description of 2d boundary vertex operator algebra (VOA) \cite{Ferrari:2023fez}. For instance, the Gang-Yamazaki (GY) minimal rank-0 SCFT \cite{Gang:2018huc} admits $M(2,5)$ minimal model and affine $osp(1|2)_1$ boundary VOAs with the A- and B-twist respectively. 

The Gang–Kim–Stubbs (GKS) $\CT_n$ theory ($n=1,2,\ldots$), a natural extension of the GY theory, was proposed in~\cite{Gang:2023rei} as a 3d $\CN=2$ abelian gauge theory that flows to a rank-0 SCFT. The boundary VOAs ontained from the A- and B-twists are $M(2n+3,2)$ and the affine $osp(1|2)_n$ VOA respectively \cite{Gang:2023rei,Ferrari:2023fez}. 

The minimal model $M(2n+3,2)$, on the other hand, also arises as the chiral algebra of the 4d $\CN=2$ $(A_1,A_{2n})$ Argyres–Douglas SCFT via SCFT/VOA correspondence~\cite{Beem:2013sza,Beem:2014cca,Beem:2014rka,Cordova:2015nma,Beem:2017ooy}, and it is expected that $\CT_n$ is obtained by the $U(1)_r$ twisted circle reduction of $(A_1,A_{2n})$ theory. Namely, $\CT_n$ bridges the 4d SCFT and the desired 2d VOA upon the topological A-twist~\cite{Dedushenko:2023cvd,ArabiArdehali:2024ysy,Gaiotto:2024ioj,ArabiArdehali:2024vli}. In the class-$\CS$ construction, the $(A_1,A_{2n})$ theory is obtained by compactifying 6d $\CN=(0,2)$ $A_1$ SCFT on a sphere with an irregular puncture~\cite{Gaiotto:2009hg,Xie:2012hs}. The $U(1)_r$ twisted circle reduction then corresponds to a fibration of the punctured sphere over the circle, resulting in a lens space~\cite{Closset:2026xjj}
\begin{align}
    L_{n,k}
    :=
    L\big(2n+3,2k\big)
    \setminus\{v\}
    \,,
    \label{eq: k lens}
\end{align}
where a vertex $v$ from the irregular puncture is removed, while $k \in \mathbb{Z}_{2n+3}^\times$ is the wrapping number of the $U(1)_r$ twist. A natural expectation, then, is that $\CT_n$ is obtained from the 3d/3d correspondence for $L_{n,1}$ as a class-$\CR$ construction~\cite{Dimofte:2011ju, Dimofte:2011py}.

\vspace{5pt}
In this paper, we {\it precisely} derive $\CT_n$ via the Dimofte-Gaiotto-Gukov (DGG) construction from the {\it minimal} triangulation of the punctured lens space $L_{n,1}$ \footnote{The $\CT_n$ theory was obtained in \cite{Gang:2025ykf} from a torus knot complement $S^3 \setminus T(2n+3,2)$ as the class-$\CR$ construction with gluing data from {\tt SnapPy} up to a procedure that reduces two tetrahedra. It will be interesting to clarify the relation between the two manifolds $L_{n,1}$ and the torus knot complement implemented by the reduction. }. Furthermore, we infinitely extend the 3d $\CN=2$ gauge theories in~\cite{Go:2025ixu} from $L_{n,n+1}$ that flow to unitary TQFTs, whose modular data lie in the same Galois orbit with that of $M(2n+3,2)$. We also conjecture self-mirror rank-0 SCFTs from the amphichiral condition of the lens space.

Before we begin, we remark on an aspect of the construction that remains unclear. The DGG construction only sees irreducible $SL(2,\mathbb{C})$ flat connections on the complex Chern-Simons theory side, loosing the reducible ones as pointed out in~\cite{Chung:2014qpa}. Since the lens space has reducible flat connections only, the partition functions of the corresponding DGG theory vanish due to spontaneously broken supersymmetry~\cite{Gang:2018gyt}. However, for the {\it punctured} lens space $L_{n,k}$, we find $\CN=2$ gauge theories with the number of supersymmetric vacua exactly matches the number of non-trivial {\it reducible} flat connections. The only difference between the two three-manifolds is the removal of a vertex $v$, which is a universal feature in the context of layered triangulations~\cite{Jaco:2006,Jaco:2009}. Thus, it would be interesting to study the role of this vertex in the DGG construction for capturing the reducible flat connections in the spirit of~\cite{Cheng:2018vpl,Gadde:2013sca,Gukov:2016gkn,Gukov:2017kmk,Eckhard:2019jgg,Chung:2019khu,Assel:2022row,Chung:2023qth,Chung:2026qfv,Chung:2026qyr}.

\subsection{Gang-Kim-Stubbs theory: 3d bulk for \texorpdfstring{$M(2,2n+3)$}{M(2n+3,2)} }
We shall start by reviewing the GKS theory $\CT_n$ labeled by $n\in \mathbb{N}$ \cite{Gang:2023rei}, which is a family of 3d $\CN=2$ $U(1)_{K}^n$ gauge theory of $n\times n$ Chern-Simons level matrix $K$
\begin{align}
    K_{ij} = 2\, \text{min}(i,j)
    \,,
    \label{eq: GKS K}
\end{align}
with $n$ chiral multiplets $\Phi_{i}$ ($i=1,\cdots,n$), charged $+\d_{i,j}$ under the $j$-th $U(1)$ factor and neutral under the R-symmetry. The superpotential is given by half-BPS monopole operators
\begin{align}
    W = \sum_{i=1}^{n-1} V_{\mathfrak{m}^{(i)}}
    \;\,\;\;
    \mathfrak{m}_j^{(i)} = 2\d_{i,j} - \d_{i,j-1} - \d_{i,j+1}
    \,,
    \label{eq: GKS superpotential}
\end{align}
where $\mathfrak{m}_j^{(i)}$ is the magnetic flux of the $j$-th $U(1)$ gauge factor. This superpotential breaks the topological symmetries except a single combination $U(1)_A$ generated by
\begin{align}
    A = \sum_{j=1}^n j \cdot T_j
    \,,
\end{align}
where $T_j$ is the generator of the $j$-th $U(1)_{T_j}$ topological symmetry. The $\CT_n$ theory flows to the $\CN=4$ rank-0 SCFT in the IR and $U(1)_A$ becomes the axial symmetry.

\medskip\noindent {\bf Topological twist}.
With $\CN=4$ supersymmetry in the IR, $\CT_n$ admits the 3d topological A-twist \cite{Kapustin:2010ag,Witten:1988ze}, or B-twist \cite{Blau:1996bx,Rozansky:1996bq}. By denoting two Cartans of $SO(4)_R \simeq SU(2)_H \times SU(2)_C$ R-symmetry by $J^H$ and $J^C$ respectively ($A=J^C - J^H$), and defining their embedding with a mixing parameter $\n$ as
\begin{align}
    R_\n := (J^C + J^H) + \n(J^C - J^H)
    \,,
\end{align}
then $\n=-1$ and $\n=+1$ realize the topological A- and B-twist respectively, while $\n=0$ corresponds to the conformal fixed point.

Since $\CT_n$ has neither Coulomb nor Higgs branch, neither twisted sector contains local operators, capturing 3d non-unitary semisimple TQFT structures. The A-/B-twisted sectors support $M(2n+3,2)$ Virasoro minimal model and affine VOA of $osp(1|2)$ at level $n$ respectively \cite{Gang:2023rei,Ferrari:2023fez}.

\subsection{Lens space \texorpdfstring{$L(p,q)$}{L(pq)}}
\begin{figure}[tbp]
\centering
\begin{tikzpicture}[scale=1.3, rotate=0]
\usetikzlibrary{calc}

\foreach \i in {1,...,9} {
    \coordinate (P\i) at ({2.3*cos(13+360/9*(\i-1))},{0.8*sin(13+360/9*(\i-1))});
}
\foreach \i in {1,...,9} {
    \coordinate (V\i) at ({2.7*cos(13+360/9*(\i-1))},{0.9*sin(13+360/9*(\i-1))});
}

\coordinate (Pn) at (0,1.7);
\coordinate (Ps) at (0,-1.7);
\node at (0,1.85) {$v_{\mathfrak{n}}$};
\node at (0,-1.85) {$v_{\mathfrak{s}}$};

\draw[line width=1pt,line join=round] (P4) -- (P5) -- (P6) -- (P7) -- (P8) -- (P9) -- (P1) ;
\draw[dotted,line width=1pt,line join=round] (P1) -- (P2) -- (P3) -- (P4) ;

\draw[line width=1pt,line join=round] (Pn) -- (P1) ;
\draw[dotted,line width=1pt,line join=round] (Pn) -- (P2) ;
\draw[dotted,line width=1pt,line join=round] (Pn) -- (P3) ;
\draw[line width=1pt,line join=round] (Pn) -- (P4) ;
\draw[line width=1pt,line join=round] (Pn) -- (P5) ;
\draw[line width=1pt,line join=round] (Pn) -- (P6) ;
\draw[line width=1pt,line join=round] (Pn) -- (P7) ;
\draw[line width=1pt,line join=round] (Pn) -- (P8) ;
\draw[line width=1pt,line join=round] (Pn) -- (P9) ;

\draw[dotted,line width=1pt,line join=round] (Ps) -- (P1) ;
\draw[dotted,line width=1pt,line join=round] (Ps) -- (P2) ;
\draw[dotted,line width=1pt,line join=round] (Ps) -- (P3) ;
\draw[dotted,line width=1pt,line join=round] (Ps) -- (P4) ;
\draw[dotted,line width=1pt,line join=round] (Ps) -- (P5) ;
\draw[line width=1pt,line join=round] (Ps) -- (P6) ;
\draw[line width=1pt,line join=round] (Ps) -- (P7) ;
\draw[line width=1pt,line join=round] (Ps) -- (P8) ;
\draw[line width=1pt,line join=round] (Ps) -- (P9) ;

\node at (V9) {$v_{p\texttt{-}2}$};
\node at (V1) {$v_{p\texttt{-}3}$};
\node at (V4) {$v_{3}$};
\node at (V5) {$v_{2}$};
\node at (V6) {$v_{1}$};
\node at (-0.98,-0.53) {$v_{0}$};
\node at (1.08,-0.45) {$v_{p\texttt{-}1}$};

\end{tikzpicture}
\caption{\label{fig: lens} The lens space $L(p,q)$ is obtained by gluing the two hemispherical $p$-gons on the boundary of a topologically 3-ball with a $q$-unit rotation.
}
\end{figure}
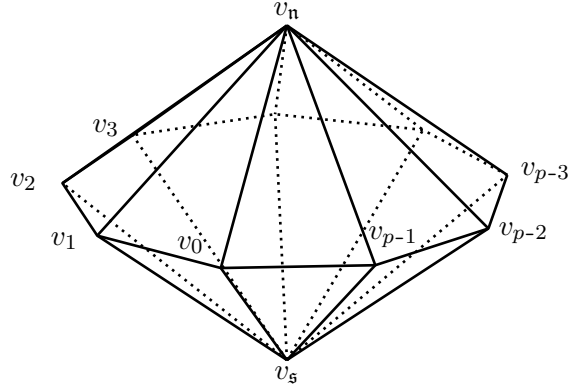
Next, we review the lens space. For a pair of positive coprime integers $p$ and $q$, a lens space $L(p,q)$ is given by a quotient of a three sphere $S^3: |z_1|^2 + |z_2|^2 = 1$ by $\mathbb{Z}_p$
\begin{align}
    L(p,q) \cong S^3 / \mathbb{Z}_p
    \;\;\;,\;\;\;\;\;
    \mathbb{Z}_p \; :\; (z_1,z_2) \sim 
    (e^{\frac{2\pi i q}{p}} z_1 , e^{\frac{2\pi i }{p}} z_2 )
    \,.
\end{align}
This can be obtained by gluing two hemispherical $p$-gons on the boundary of a 3-ball as depicted in Figure \ref{fig: lens} where $v_{\mathfrak{n}}$ and $v_{\mathfrak{s}}$ denote the north and south poles respectively, and $v_i$ ($i=0,\cdots,p-1$) are the vertices of the $p$-gons. The identification is performed with a $q$-unit rotation: the 2-simplex $(v_\mathfrak{n} v_{[i]_p} v_{[i+1]_p})$ is identified with $(v_{\mathfrak{s}} v_{[i+q]_p} v_{[i+1+q]_p} )$ for each $i$.

The lens space can also be described as a Seifert manifold with a single orbifold point
\begin{align}
    L(p,q) \cong [ d=0; g=0; (q,p) ]
    \,,
\end{align}
where $d$ is the degree of the Seifert fibration and $g$ is the genus of the base, thus $S^2$ in this case. A disk neighborhood near the orbifold point on $S^2$ forms a solid torus
\begin{align}
    T(p,t) \cong D^2 \times_{t/p} S^1
    \;\;,\;\;\;
    qt =1\, \text{mod}\, p
    \,,
\end{align}
under the fibration whose angle coordinates $(\varphi, \psi)$ satisfy
\begin{align}
    (\varphi, \psi) \sim (\varphi + 2\pi, \psi + \frac{2\pi t}{p})
    \,,
    \label{eq: disk identification}
\end{align}
where $\varphi$ and $\psi$ describe the contractible and non-contractible cycles, respectively. 

In this paper, we focus on $p=2n+3$ to describe the $(2n+3)$-gon, that is the Gaiotto curve of the $(A_1,A_{2n})$ theory~\cite{Gaiotto:2009hg,Cecotti:2011iy}, and $t=2k$ to implement a 2-unit rotation of the polygon per each $U(1)_r$ twist. Thus, with a homeomorphism $L(p,q) \cong L(p,q^{-1})$, we get
\begin{align}
    L(2n+3,(2k)^{-1}) \cong L(2n+3,2k)
    \,.
\end{align}

\section{Dimofte-Gaiotto-Gukov construction of punctured lens space \texorpdfstring{$L_{n,k}$}{Lnk}} \label{sec: derivation}
In this section, we demonstrate that the punctured lens space $L_{n,1}$ gives rise to the GKS $\CT_n$ theory via the DGG construction. We also construct a 3d $\CN=2$ abelian gauge theory from $L_{n,n+1}$ that flows to a 3d unitary TQFT in the IR whose modular data lies in the Galois orbit with that of $M(2n+3,2)$ minimal model. Moreover, we conjecture a condition for the resulting DGG theory being a self-mirror rank-0 SCFT.

\subsection{Review on Dimofte-Gaiotto-Gukov construction}
The DGG theory $T[M]$ is a 3d $\CN=2$ gauge theory obtained by compactifying 6d $\CN=(0,2)$ $A_1$ SCFT on a 3-manifold $M$ with topological twist by $SO(3)_R \in SO(5)_R$ R-symmetry \cite{Dimofte:2011ju}. Consider an ideal triangulation of $M$ into $N$ ideal tetrahedra
\begin{align}
    M = \bigcup_{i=1}^N \D_i
    \,,
    \label{eq: triangulation}
\end{align}
where $\D_i$ denotes the $i$-th ideal tetrahedron with hyperbolic structure determined by three complex variables $(Z_i,Z_i',Z_i'')$, defining the boundary phase space~\cite{Dimofte:2011gm}
\begin{align}
    \CP_{\partial\D_i}
    =
    \Big\{
    Z_i,Z_i',Z_i'' \in \mathbb{C}\backslash 2\pi i \mathbb{Z}
    \,
    \Big|\,
    Z_i + Z_i' + Z_i'' = \pi i
    \Big\}\,,
\end{align}
with Poisson brackets $\{ Z_i,Z_j' \}=\{ Z_i',Z_j'' \}=\{Z_i'',Z_j \} =\d_{i,j}$. There are three possible polarization choices $\s_i$ of canonically conjugate coordinates $(X_i,P_i)$ for each $\CP_{\partial \D_i}$
\begin{align}
    (X_i,P_i)
    =
    \Bigg\{
    \begin{array}{cc}
        (Z_i,Z_i'') & \;\;\text{if}\;\; \s_i = 0 \\
        (Z_i'',Z_i') & \;\;\text{if}\;\; \s_i = 1 \\
        (Z_i',Z_i) & \;\,\;\;\text{if}\;\; \s_i = 2\,.
    \end{array}
\end{align}
Upon gluing back the triangulation \eqref{eq: triangulation} with polarization choice $\s_i = 0$, the faces of the tetrahedra are identified pairwise, resulting in $N$ linearly independent edge variables
\begin{align}
    C_i := 
    \sum_{i=1}^N 
    \Big(
    g_{ij}^{(0)} Z_j
    +
    g_{ij}^{(1)} Z_j'
    +
    g_{ij}^{(2)} Z_j''
    \Big)
    \,,
\end{align}
that mutually commute $\{ C_i,C_j \} =0 $ with coefficients $g_{ij}^{(s)} \in \{0,1,2\}$. If $C_I$ is an internal edge, it satisfies $C_{I\in \text{internal}} = 2\pi i$ for smooth metric on $M$, and the boundary phase space of $M$ becomes a symplectic quotient of the product of the tetrahedra boundary phase spaces
\begin{align}
    \CP_{\partial M} = 
    \bigg(
    \prod_{i=1}^N \CP_{\partial \D_i}
    \bigg)
    \Big\slash\!\!\!\Big\slash
    \{C_{I\in \text{internal}} = 2\pi i\}
    \,.
\end{align}
One can similarly find $N$ linearly independent conjugate variables $\G_i$ that satisfy
\begin{align}
    \{C_i,\G_j\} = - \d_{i,j} 
    \;\;,\;\;\;
    \{\G_i,\G_j\} = 0
    \,,
\end{align}
so that an element $h \in Sp(2N,\mathbb{Z})$ is determined by the relation
\begin{align}
    \left(
    \begin{array}{c}
        C_1 \\
        \vdots \\
        C_N \\
        \hline
        \G_1 \\
        \vdots \\
        \G_N
    \end{array}
    \right)
    =
    h \cdot
    \left(
    \begin{array}{c}
        X_1 \\
        \vdots \\
        X_N \\
        \hline
        P_1 \\
        \vdots \\
        P_N
    \end{array}
    \right)
    \,,
\end{align}
which can be decomposed into three generators
\begin{align}
    h_S = \left(
    \begin{array}{c|c}
        {\bf 1}-\L & - \L  \\
         \hline
        \L & {\bf 1}-\L
    \end{array}
    \right)
    \;,\;\;
    h_T = \left(
    \begin{array}{c|c}
        {\bf 1} & {\bf 0} \\
         \hline
         \CK & {\bf 1}
    \end{array}
    \right)
    \;,\;\;
    h_U = \left(
    \begin{array}{c|c}
        U & {\bf 0} \\
         \hline
         {\bf 0} & U^{-1t}
    \end{array}
    \right)
    \,,
    \label{eq: decomposition of h}
\end{align}
where $\L = \text{diag}(\l_1,\cdots,\l_N)$ with $\l_i \in \{0,1\}$, and $U \in GL(N,\mathbb{Z})$. 

Each tetrahedron $\D_i$ corresponds to a 3d $\CN=2$ theory $T_{\D_i}$ of a chiral multiplet coupled to a background $U(1)$ flavor symmetry with effective CS level $-1/2$. Then, $h$ encodes a 3d $\CN=2$ gauge theory as an action on the $N$ free chiral multiplets $\{T_{\D_i}\}$ according to the decomposition \eqref{eq: decomposition of h} where $h_U$ shuffles the $U(1)^N$ background gauge fields by $U$, $h_T$ introduces mixed CS terms with level matrix $\CK$, and $h_S$ makes the $i$-th background gauge field dynamical if $\l_i=1$. 

The superpotential of the DGG theory consists of operators $O_I$ from the internal edges of {\it easy} type defined as {\it at most one of} $(g_{Ij}^{(0)},g_{Ij}^{(1)},g_{Ij}^{(2)})$ {\it is non-zero for each} $j$:
\begin{align}
    W = \sum_{C_I \in \text{easy}} O_I
    \,.
    \label{eq: easy potential}
\end{align}
Therefore, we would schematically write the 3d $\CN=2$ gauge theory as
\begin{align}
    T\big[M,\{\s_i=0\}\big]
    =
    h \circ 
    \bigg(
    \bigotimes_{i=1}^N T_{\D_i}
    \bigg)
    \;\; \text{with superpotential \eqref{eq: easy potential}}
    \,.
\end{align}
For generic polarization choices $\s = \{\s_i\}$, the edge variable coefficients simply shuffle as
\begin{align}
    g_{ij} = g_{ij}^{([\s_j])}
    \;,\;\;
    g'_{ij} = g_{ij}^{([\s_j+1])}
    \;,\;\;
    g''_{ij} = g_{ij}^{([\s_j+2])}
    \,,
\end{align}
with $[s]:= s\,\text{mod}\,3$. The gauge theory descriptions $T[M,\s]$ are distinct for different polarizations; nevertheless, they flow to the same IR fixed point which depends only on the topology of $M$, thus we denote the fixed point by $T[M]$.

\medskip\noindent {\bf Simple gauge theory description}.
The two matrices $(A,B)$ of $ h = \scalebox{0.6}{$\begin{pmatrix} A & B \\ C & D \end{pmatrix}$}\in Sp(2N,\mathbb{Z})$ are called {\it Neumann-Zagier} (NZ) matrices \cite{NeumannZagier1985,Dimofte:2012qj}, and it often is possible to choose a polarization $\s$ such that $|\text{det}(B)|=1$. This simplifies the UV description of $T[M]$ \cite{Dimofte:2012qj,Gang:2025ykf} as 3d $\CN=2$ $U(1)^N$ gauge theory with $N$ chirals $\Phi_i$ of charge $+\d_{i,j}$ under the $j$-th $U(1)$ and vanishing R-charge. The CS level matrix becomes
\begin{align}
    K = B^{-1}A
    \,,
\end{align}
and the superpotential term $O_I$ for each easy internal edge is given by
\begin{align}
    O_I = \Big( \prod_{j=1}^{N}\phi_j^{g_{Ij}} \Big) V_{\mathfrak{m}^{(I)}}
    \;,\;\;
    \mathfrak{m}_j^{(I)} = -B_{Ij}
    \,,
    \label{eq: monopole term}
\end{align}
where $\phi_j$ is the scalar component of $\Phi_j$ and $V_{\mathfrak{m}^{(I)}}$ is the bare monopole operator.

\medskip\noindent {\bf Orientation reversal}.
Suppose we have a DGG theory $T[M,\s]$ from an ideal triangulation of an oriented 3-manifold $M$ with polarization $\s$ that admits $|\text{det}(B)|=1$. Then, the DGG theory $T[\widetilde{M},\s]$ of the orientation reversal of $M$ is simply established by reversing all the orientations of the ideal tetrahedra, say as $Z_i' \leftrightarrow Z_i''$, which affects
\begin{align}
    \Wg = g
    \;,\;\;
    \Wg' = g''
    \;,\;\;
    \Wg'' = g'
    \,,
\end{align}
for $\Wg$, $\Wg'$, and $\Wg''$ are the coefficient matrices from the triangulation of $\widetilde{M}$. The corresponding NZ matrices then become
\begin{align}
    \widetilde{A} = A - B
    \;,\;\;
    \widetilde{B} = - B
    \,,
\end{align}
which result in CS level matrix $\WK$ and superpotential terms $\WO_I$ of $T[\widetilde{M},\s]$ as
\begin{align}
    \WK = {\bf 1} - K
    \;,\;\;
    \WO_I = O_I |_{\mathfrak{m}^{(I)}\to -\mathfrak{m}^{(I)}}
    \,,
    \label{eq: parity conj}
\end{align}
meaning the UV gauge theory $T[\widetilde{M},\s]$ is the 3d parity conjugation of $T[M,\s]$.

\subsection{Minimal triangulation of \texorpdfstring{$L_{n,1}$}{Ln1}: Gang-Kim-Stubbs \texorpdfstring{$\CT_n$}{Tn}}
\begin{figure}[tbp]
\centering
\begin{tikzpicture}[scale=2.2, rotate=90]
\usetikzlibrary{calc}

%RHS%
\foreach \i in {1,...,11} {
    \coordinate (P\i) at ({cos(360/11*(\i-1))},{sin(360/11*(\i-1))});
}
\foreach \i in {1,...,11} {
    \coordinate (V\i) at ({1.1*cos(360/11*(\i-1))},{1.12*sin(360/11*(\i-1))});
}

\node at (V1) { $\cdots$};
\node at (V2) {\tiny $v_1$};
\node at (V3) {\tiny $v_0$};
\node at (V4) {\tiny $v_{2n\texttt{+}2}\;\;$};
\node at (V5) {\tiny $v_{2n\texttt{+}1}$};
\node at (V6) {\tiny $v_{2n}$};
\node at (-1.05,0) {$\cdots$};
\node at (V7) {\tiny $v_{n\texttt{+}3}$};
\node at (V8) {\tiny $v_{n\texttt{+}2}$};
\node at (V9) {\tiny $\;\;v_{n\texttt{+}1}$};
\node at (V10) {\tiny $v_{n}$};
\node at (V11) {\tiny $v_{n\texttt{-}1}$};

\draw[line width=1pt,line join=round,blue] (P3) -- (P5) ;
\draw[line width=1pt,line join=round,blue] (P2) -- (P6) ;
\draw[line width=1pt,line join=round,blue] (P1) -- (P7) ;
\draw[line width=1pt,line join=round,blue] (P11) -- (P8) ;
\draw[line width=1pt,line join=round,red] (P3) -- (P6) ;
\draw[line width=1pt,line join=round,red] (P2) -- (P7) ;
\draw[line width=1pt,line join=round,red] (P1) -- (P8) ;
\draw[line width=1pt,line join=round,red] (P11) -- (P9) ;

\draw[line width=1pt] (P1)
\foreach \i in {2,...,11} { -- (P\i) } -- cycle;

\node[rotate=18] at (0.35,0.1) {$\cdots$};

\node[rotate=-80] at (-0.15,0.89) {\scalebox{1}{\tiny $b_{2n\texttt{+}1}$}};
\node at (0.35,-0.86) {\tiny $b_{1}$};

\node at (-0.3,0.7) {\tiny $b_{2n}$};
\node at (0.3,-0.5) {\tiny $b_{3}$};

\node at (0.15,0.63) {\tiny $b_{2n\texttt{-}1}$};
\node at (0.,-0.8) {\tiny $b_{2}$};

\node at (-0.4,0.17) {\tiny $b_{2n\texttt{-}2}$};
\node at (-0.35,-0.4) {\tiny $b_{4}$};

%LHS%
\begin{scope}[shift={(0,3.1)}]
\foreach \i in {1,...,11} {
    \coordinate (P\i) at ({cos(360/11*(\i-1))},{sin(360/11*(\i-1))});
}
\foreach \i in {1,...,11} {
    \coordinate (V\i) at ({1.1*cos(360/11*(\i-1))},{1.12*sin(360/11*(\i-1))});
}

\draw[line width=1pt,line join=round,blue] (P3) -- (P5) -- (P2) -- (P6) -- (P1) -- (P7) -- (P11) -- (P8) -- (P10);

\node at (-0.3,0) {$\cdots$};

\node at (-0.15,0.9) {\tiny $a_1$};
\node[rotate=80] at (-0.14,-0.9) {\scalebox{1}{\tiny $a_{2n\texttt{+}1}$}};

\node at (-0.15,0.53) {\tiny $a_{3}$};
\node at (-0.24,-0.54) {\tiny $a_{2n\texttt{-}1}$};

\node at (0.15,0.75) {\tiny $a_{2}$};
\node at (0.15,-0.76) {\tiny $a_{2n}$};

\node at (0.3,0.27) {\tiny $a_{4}$};
\node at (0.3,-0.28) {\tiny $a_{2n\texttt{-}2}$};

\node at (V1) { $\cdots$};
\node at (V2) {\tiny $v_1$};
\node at (V3) {\tiny $v_0$};
\node at (V4) {\tiny $v_{2n\texttt{+}2}\;\;$};
\node at (V5) {\tiny $v_{2n\texttt{+}1}$};
\node at (V6) {\tiny $v_{2n}$};
\node at (-1.05,0) {$\cdots$};
\node at (V7) {\tiny $v_{n\texttt{+}3}$};
\node at (V8) {\tiny $v_{n\texttt{+}2}$};
\node at (V9) {\tiny $\;\;v_{n\texttt{+}1}$};
\node at (V10) {\tiny $v_{n}$};
\node at (V11) {\tiny $v_{n\texttt{-}1}$};

\draw[line width=1pt] (P1)
\foreach \i in {2,...,11} { -- (P\i) } -- cycle;

\end{scope}

\node at (0,1.6) {\scriptsize Flip};
\node at (-0.2,1.6) {$\longrightarrow$};

\end{tikzpicture}

\caption{\label{fig: flip} Ideal triangulations of $(2n+3)$-gon. In the left, we triangulate it with zigzag pattern to get $2n+1$ triangles denoted by $a_i$. By flipping the edges $\overline{v_{i} v_{2n-i}}$ for $i=0,\cdots, n-1$, we obtain the same triangulation rotated counterclockwise by $(n+2)$-unit, with triangles denoted by $b_i$.
}
\end{figure}
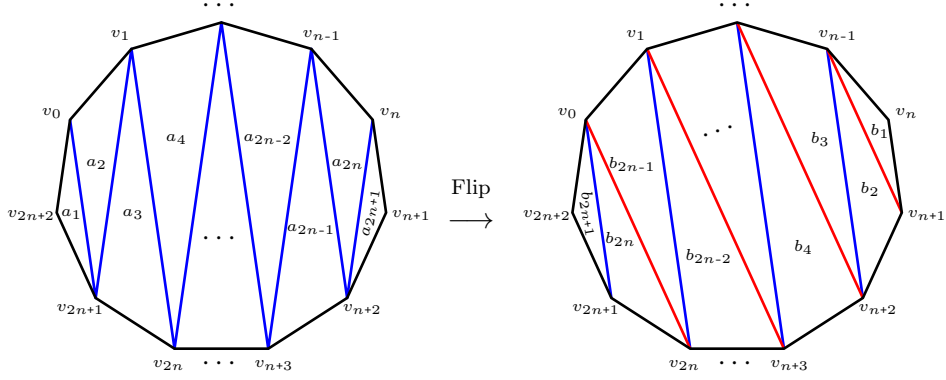
Consider a $(2n+3)$-gon, which is the Gaiotto curve of the $(A_1,A_{2n})$ theory, triangulated into $2n+1$ triangles as in the left of Figure \ref{fig: flip} where we denote each triangle by $a_i$. We also label the vertices as $v_0,\cdots,v_{2n+2}$. By flipping $n$ edges $\overline{v_{i} v_{2n-i}}$ for $i=0,\cdots, n-1$ as shown in red, we get another triangulation of the same polygon in the right hand side where each triangle is denoted by $b_i$.

\medskip\noindent {\bf From polygon to lens space}.
Note that the triangulation of the polygon in the right of Figure \ref{fig: flip} is the same as the left one by anticlockwise $(n+2)$-unit rotation. If we view the left and right polygons as the bottom and top faces of a polyhedron in Figure \ref{fig: lens}, the pairwise identification $a_i \sim b_i$ for $i=1,\cdots,2n+1$ produces $L(2n+3,n+2)\cong L(2n+3,2)$. This realizes a fibration of the Gaiotto curve over the circle corresponding to once wrapping $U(1)_r$ twisted circle reduction of the $(A_1,A_{2n})$ theory~\cite{Closset:2026xjj}.

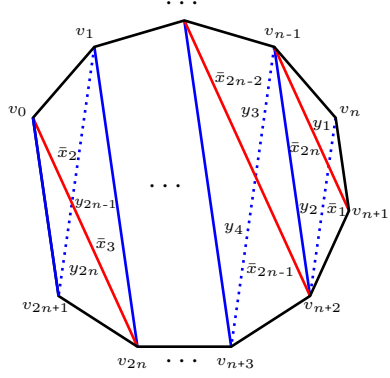
\begin{figure}[tbp]
\centering
\begin{tikzpicture}[scale=2.2, rotate=90]
\usetikzlibrary{calc}

%RHS%
\foreach \i in {1,...,11} {
    \coordinate (P\i) at ({cos(360/11*(\i-1))},{sin(360/11*(\i-1))});
}
\foreach \i in {1,...,11} {
    \coordinate (V\i) at ({1.1*cos(360/11*(\i-1))},{1.1*sin(360/11*(\i-1))});
}

\node at (V1) { $\cdots$};
\node at (V2) {\tiny $v_1$};
\node at (V3) {\tiny $v_0$};
\node at (V5) {\tiny $v_{2n\texttt{+}1}$};
\node at (V6) {\tiny $v_{2n}$};
\node at (-1.05,0) {$\cdots$};
\node at (V7) {\tiny $v_{n\texttt{+}3}$};
\node at (V8) {\tiny $v_{n\texttt{+}2}$};
\node at (V9) {\tiny $\;\;v_{n\texttt{+}1}$};
\node at (V10) {\tiny $v_{n}$};
\node at (V11) {\tiny $v_{n\texttt{-}1}$};

\draw[dotted,line width=1pt,line join=round,blue] (P2) -- (P5) ;
\draw[dotted,line width=1pt,line join=round,blue] (P11) -- (P7) ;
\draw[dotted,line width=1pt,line join=round,blue] (P10) -- (P8) ;
\draw[line width=1pt,line join=round,blue] (P2) -- (P6) ;
\draw[line width=1pt,line join=round,blue] (P1) -- (P7) ;
\draw[line width=1pt,line join=round,blue] (P11) -- (P8) ;
\draw[line width=1pt,line join=round,red] (P3) -- (P6) ;
\draw[line width=1pt,line join=round,red] (P1) -- (P8) ;
\draw[line width=1pt,line join=round,red] (P11) -- (P9) ;
\draw[line width=1pt] (P1)
\foreach \i in {2,3,5,6,7,8,9,10,11} { -- (P\i) } -- cycle;
\draw[line width=1pt,line join=round,blue] (P3) -- (P5) ;

\node at (0,0.1) {$\cdots$};

\node at (0.37,-0.83) {\tiny $y_{1}$};
\node at (0.23,-0.73) {\tiny $\bar{x}_{2n}$};
\node at (-0.13,-0.75) {\tiny $y_{2}$};
\node at (-0.14,-0.92) {\tiny $\bar{x}_{1}$};

\node at (0.45,-0.4) {\tiny $y_{3}$};
\node at (0.65,-0.33) {\tiny $\bar{x}_{2n\texttt{-}2}$};
\node at (-0.25,-0.3) {\tiny $y_{4}$};
\node at (-0.5,-0.52) {\tiny $\bar{x}_{2n\texttt{-}1}$};

\node at (0.2,0.7) {\tiny $\bar{x}_{2}$};
\node at (-0.5,0.6) {\tiny $y_{2n}$};
\node at (-0.35,0.47) {\tiny $\bar{x}_{3}$};
\node at (-0.1,0.53) {\scalebox{0.9}{\tiny $y_{2n\texttt{-}1}$}};

\end{tikzpicture}
\caption{\label{fig: tri1} An ideal triangulation of the punctured lens space $L_{n,1}$ into $n$ ideal tetrahedra by interpreting each flip in Figure \ref{fig: flip} as introducing a tetrahedron. The triangle faces on the bottom and top sides are denoted by $\bar{x}_i$ and $y_i$ respectively.
}
\end{figure}

\medskip\noindent {\bf Flip and tetrahedron}.
Indeed, for each {\it flip}, we introduce an {\it ideal tetrahedron} --- that is the four vertices of a tetrahedron are truncated --- such that two diagonal edges of the tetrahedron are identified to the edges before and after the flip~\cite{Cecotti:2011iy}. Then, the $n$ flips in Figure \ref{fig: flip} produce $n$ ideal tetrahedra as in Figure \ref{fig: tri1} where the bottom and top faces of each ideal tetrahedron are labeled by $\bar{x}_i$ and $y_i$ respectively for $i=1,\cdots,2n$. As we will explain soon, a pairwise gluing of these faces gives rise to the punctured lens space $L_{n,1}$.

The triangle $y_i$ is naturally mapped to $b_i$ in the right polygon in Figure \ref{fig: flip}, which is identified to $a_i$ to implement $L(2n+3,2)$. For $i\neq 1$, $a_i$ is mapped to a bottom face $\bar{x}_i$ so that we get an identification $y_i \sim x_i$. For $i=1$, $a_1$ is not directly mapped to $x_1$, but to $b_{2n+1}$ which is identified with $a_{2n+1}$, then to the bottom face $\bar{x}_1$. Consequently, the face gluing is summarized as
\begin{align}
    y_i \sim x_i
    \,,
    \label{eq: pairwise face gluing}
\end{align}
for $i=1,\cdots,2n$. In terms of ordered 2-simplex, this can be precisely given as
\begin{align}
    &y_1 = (v_{n-1} v_{n+1} v_{n})
    \sim
    (v_{n} v_{n+2} v_{n+1}) = x_1
    \,,
    \nonumber\\
    &y_{2j-1} = (v_{n-j} v_{n+j} v_{n-j+1}) 
    \sim
    (v_{2n-j+2} v_{j-1} v_{2n-j+3}) = x_{2j-1}
    \;,\;\;
    \text{for}\;\; j=2,\cdots,n
    \nonumber\\
    &y_{2j} = (v_{n+1} v_{n-j} v_{n+j+1}) \sim (v_{j-1} v_{2n-j+2} v_{j}) = x_{2j}
    \;,\;\;
    \text{for}\;\; j=1,\cdots,n
    \,.
\end{align}

\medskip\noindent {\bf The puncture}.
For each ideal tetrahedron, 4 small triangle surfaces arise from the tip truncation. After the gluing \eqref{eq: pairwise face gluing}, these small $4n$ triangles form an $S^2$ boundary inside the resulting 3-manifold. We denote a small 3-ball by $v$ that might fill in this $S^2$ hole. In general, every lens space admits a layered triangulation with one vertex that corresponds to $v$~\cite{Jaco:2006,Jaco:2009} and we would write the punctured lens space as $L(p,q)\setminus \{v\}$.

\medskip\noindent {\bf Minimal triangulation}.
The minimal triangulation of $L(p,q)\setminus \{v\}$ is conjectured \cite{Matveev:1990,Jaco:2006,Jaco:2009}
\begin{align}
    \text{min}\,\D\big(L(p,q)\big) = \sum_{i=0}^{} c_i - 3
    \;,\;\;
    \text{for}\; p>3
    \,,
    \label{eq: min conj}
\end{align}
where $c_i\in \mathbb{N}$ are the partial denominators of a continued fraction
\begin{align}
    \frac{p}{q}
    =
    c_0 + \frac{1}{c_1 + \frac{1}{c_2 + \frac{1}{c_3 + \cdots}}     }
    \,.
\end{align}
For $p=2n+3$ and $q=2$, we get $c_0 = n+1$ and $c_1 = 2$ so that the minimal triangulation has $n$ tetrahedra, based on which, our triangulation of $L_{n,1}$ in Figure \ref{fig: tri1} is the minimal one.

\begin{figure}[tbp]
\centering
\begin{tikzpicture}[scale=1.15, rotate=0]
\usetikzlibrary{calc}

\coordinate (V0) at (0,3);
\coordinate (V1) at (1,3);
\coordinate (V2) at (2,3);
\coordinate (V3) at (3,3);
\coordinate (V4) at (4.5,3);
\coordinate (V5) at (5.5,3);
\coordinate (V6) at (7,3);
\coordinate (V7) at (8,3);
\coordinate (V8) at (9,3);
\coordinate (V9) at (10,3);

\coordinate (v0) at (0,3+0.2);
\coordinate (v1) at (1,3+0.2);
\coordinate (v2) at (2,3+0.2);
\coordinate (v3) at (3,3+0.2);
\coordinate (v4) at (4.5,3+0.2);
\coordinate (v5) at (5.5,3+0.2);
\coordinate (v6) at (7,3+0.2);
\coordinate (v7) at (8,3+0.2);
\coordinate (v8) at (9,3+0.2);
\coordinate (v9) at (10,3+0.2);

\coordinate (B0) at (0,0);
\coordinate (B1) at (1,0);
\coordinate (B2) at (2,0);
\coordinate (B3) at (3,0);
\coordinate (B4) at (4.5,0);
\coordinate (B5) at (5.5,0);
\coordinate (B6) at (7,0);
\coordinate (B7) at (8,0);
\coordinate (B8) at (9,0);
\coordinate (B9) at (10,0);

\coordinate (b0) at (0,0-0.2);
\coordinate (b1) at (1,0-0.2);
\coordinate (b2) at (2,0-0.2);
\coordinate (b3) at (3,0-0.2);
\coordinate (b4) at (4.5,0-0.2);
\coordinate (b5) at (5.5,0-0.2);
\coordinate (b6) at (7,0-0.2);
\coordinate (b7) at (8,0-0.2);
\coordinate (b8) at (9,0-0.2);
\coordinate (b9) at (10,0-0.2);

\draw[dotted,blue,line width=1pt,line join=round] (V1) -- (B0) ;
\draw[dotted,blue,line width=1pt,line join=round] (V2) -- (B1) ;
\draw[dotted,blue,line width=1pt,line join=round] (V3) -- (B2) ;
\draw[dotted,blue,line width=1pt,line join=round] (V5) -- (B4) ;
\draw[dotted,blue,line width=1pt,line join=round] (V7) -- (B6) ;
\draw[dotted,blue,line width=1pt,line join=round] (V8) -- (B7) ;
\draw[dotted,blue,line width=1pt,line join=round] (V9) -- (B8) ;

\draw[red,line width=1pt,line join=round] (V0) -- (B1) ;
\draw[red,line width=1pt,line join=round] (V1) -- (B2) ;
\draw[red,line width=1pt,line join=round] (V2) -- (B3) ;
\draw[red,line width=1pt,line join=round] (V4) -- (B5) ;
\draw[red,line width=1pt,line join=round] (V6) -- (B7) ;
\draw[red,line width=1pt,line join=round] (V7) -- (B8) ;
\draw[red,line width=1pt,line join=round] (V8) -- (B9) ;

\draw[line width=1pt,line join=round] (V0) -- (V9) -- (B9) -- (B0) ;
\draw[blue,line width=1pt,line join=round] (V0) -- (B0) ;
\draw[blue,line width=1pt,line join=round] (V1) -- (B1) ;
\draw[blue,line width=1pt,line join=round] (V2) -- (B2) ;
\draw[blue,line width=1pt,line join=round] (V3) -- (B3) ;
\draw[blue,line width=1pt,line join=round] (V4) -- (B4) ;
\draw[blue,line width=1pt,line join=round] (V5) -- (B5) ;
\draw[blue,line width=1pt,line join=round] (V6) -- (B6) ;
\draw[blue,line width=1pt,line join=round] (V7) -- (B7) ;
\draw[blue,line width=1pt,line join=round] (V8) -- (B8) ;

\node at (3.75,1.5) {$\cdots$};
\node at (6.25,1.5) {$\cdots$};

\node at (v0) {\tiny $v_0$};
\node at (v1) {\tiny $v_1$};
\node at (v2) {\tiny $v_2$};
\node at (v3) {\tiny $v_3$};
\node at (v4) {\tiny $v_{\r(n)}$};
\node at (v5) {\tiny $v_{\r(n)\texttt{+}1}$};
\node at (v6) {\tiny $v_{n\texttt{-}3}$};
\node at (v7) {\tiny $v_{n\texttt{-}2}$};
\node at (v8) {\tiny $v_{n\texttt{-}1}$};
\node at (v9) {\tiny $v_{n}$};

\node at (b0) {\tiny $v_{2n\texttt{+}1}$};
\node at (b1) {\tiny $v_{2n}$};
\node at (b2) {\tiny $v_{2n\texttt{-}1}$};
\node at (b3) {\tiny $v_{2n\texttt{-}2}$};
\node at (b4) {\tiny $v_{2n\texttt{+}1\texttt{-}\r(n)}$};
\node at (b5) {\tiny $v_{2n\texttt{-}\r(n)}$};
\node at (b6) {\tiny $v_{n\texttt{+}4}$};
\node at (b7) {\tiny $v_{n\texttt{+}3}$};
\node at (b8) {\tiny $v_{n\texttt{+}2}$};
\node at (b9) {\tiny $v_{n\texttt{+}1}$};

\node at (0+0.15,1.5) {\tiny $2''$};
\node at (1+0.15,1.5) {\tiny $4''$};
\node at (2+0.15,1.5) {\tiny $6''$};
\node at (4.5+0.15,1.5) {\tiny $n''$};
\node at (7+0.15,1.5) {\tiny $5''$};
\node at (8+0.15,1.5) {\tiny $3''$};
\node at (9+0.15,1.5) {\tiny $1''$};

\node at (1-0.15,1.5) {\tiny $2''$};
\node at (2-0.15,1.5) {\tiny $4''$};
\node at (3-0.15,1.5) {\tiny $6''$};
\node at (5.5-0.15,1.5) {\tiny $n''$};
\node at (8-0.15,1.5) {\tiny $5''$};
\node at (9-0.15,1.5) {\tiny $3''$};
\node at (10-0.15,1.5) {\tiny $1''$};

\node at (0.5,3-0.15) {\tiny $2$};
\node at (0.5,0+0.15) {\tiny $2$};
\node at (1.5,3-0.15) {\tiny $4$};
\node at (1.5,0+0.15) {\tiny $4$};
\node at (2.5,3-0.15) {\tiny $6$};
\node at (2.5,0+0.15) {\tiny $6$};
\node at (5,3-0.15) {\tiny $n$};
\node at (5,0+0.15) {\tiny $n$};
\node at (7.5,3-0.15) {\tiny $5$};
\node at (7.5,0+0.15) {\tiny $5$};
\node at (8.5,3-0.15) {\tiny $3$};
\node at (8.5,0+0.15) {\tiny $3$};
\node at (9.5,3-0.15) {\tiny $1$};
\node at (9.5,0+0.15) {\tiny $1$};

\node at (0.5+0.02,1.85) {\tiny $2'$};
\node at (1.5+0.02,1.85) {\tiny $4'$};
\node at (2.5+0.02,1.85) {\tiny $6'$};
\node at (5+0.02,1.85) {\tiny $n'$};
\node at (7.5+0.02,1.85) {\tiny $5'$};
\node at (8.5+0.02,1.85) {\tiny $3'$};
\node at (9.5+0.02,1.85) {\tiny $1'$};

\end{tikzpicture}
\caption{\label{fig: tri2} Minimal ideal triangulation of the punctured lens space $L_{n,1}$ into $n$ ideal tetrahedra. We denote $(i,i',i'')$ for the edge variables $(Z_i,Z'_i,Z''_i)$ of the $i$-th ideal tetrahedron. We also denote the vertices as $v_j$ with $\r(n)=\lfloor \frac{n-1}{2} \rfloor$.
}
\end{figure}
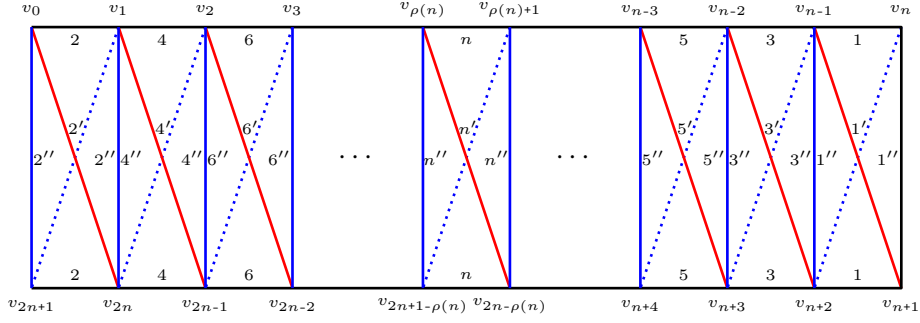

\medskip\noindent {\bf The DGG theory}.
Let us reorganize the $n$ ideal tetrahedra of $L_{n,1}$ in Figure \ref{fig: tri1} as in Figure \ref{fig: tri2} where we numbered the tetrahedra in an alternating manner for later purpose. From the gluing \eqref{eq: pairwise face gluing}, we find the $n$ independent edge variables
\begin{align}
    C_i = 2 Z_i' + (1-\d_{i,1})Z_{i-1}'' +  (1-\d_{i,n})Z_{i+1}'' + \d_{i,n} Z_n''
    \,,
\end{align}
for $i=1,\cdots, n$. With polarization choices $\s_i = 0$, the coefficients of them are organized as
\begin{align}
    g = {\bf 0}_{n\times n}
    \;,\;\;\;\;\;\;\;
    g' = 2 \cdot {\bf 1}_{n\times n}
    \;,\;\;\;\;\;\;\;
    g'' = 
    \left(
    \begin{array}{ccccccc}
        0\; &\; 1 &  & &  &  &  \\
        1 \;& 0 \; & 1\; & &  & &  \\
         & 1 & 0 & 1 & &  & \\
          &  & 1 &  \ddots & \ddots &  & \\
          & & & \ddots & 0 & 1 & \\
         & &  &  & 1 & 0 \;& 1 \\
         &  & &  & & 1 \; & 1
    \end{array}
    \right)
    \,,
\end{align}
from which, we get the Neumann-Zagier matrices $A = g-g'$ and $B=g''-g'$
\begin{align}
    A = -2 \cdot {\bf 1}_{n\times n} 
    \;,\;\;\;\;\;\;
    B = 
    \left(
    \begin{array}{ccccccc}
        -2 & 1 &  & &  &  &  \\
        1 & -2 & 1 & &  & &  \\
         & 1 & -2 & 1 & &  & \\
          &  & 1 &  \ddots & \ddots &  & \\
          & & & \ddots & -2 & 1 & \\
         & &  &  & 1 & -2 & 1 \\
         &  & &  & & 1 & -1
    \end{array}
    \right)
    \,,
\end{align}
or, explicitly $B_{ij}= -2 \d_{i,j} + \d_{i,j-1} + \d_{i,j+1}$.

Since $\text{det}(B)=(-1)^n$, the DGG theory $T[ L_{n,1},\s]$ has a simple description as a 3d $\CN=2$ $U(1)^n$ gauge theory with $n$ chiral multiplets $\Phi_i$ ($i=1,\cdots,n$) charged $+\d_{i,j}$ under the $j$-th $U(1)$ and neutral under the R-symmetry. The CS level matrix $K = B^{-1}A$ reads
\begin{align}
    K_{ij} = 2\, \text{min}(i,j)\,,
\end{align}
and the superpotential consists of $n-1$ terms from easy internal edges $C_i$ ($i=1,\cdots,n-1$)
\begin{align}
    W = \sum_{i=1}^{n-1} V_{\mathfrak{m}^{(i)}}
    \;\;,
    \;\;\;
    \mathfrak{m}_j^{(i)} = 2 \d_{i,j} - \d_{i,j-1} - \d_{i,j+1}
    \,,
\end{align}
where $K$ and $W$ are the ones of $\CT_n$ theory. Therefore, the DGG theory of the minimal triangulation of $L_{n,1}$ precisely produces the GKS $\CT_n$ theory.

\subsection{Minimal triangulation of \texorpdfstring{$L_{n,n+1}$}{Ln,n+1}: unitary TQFT}
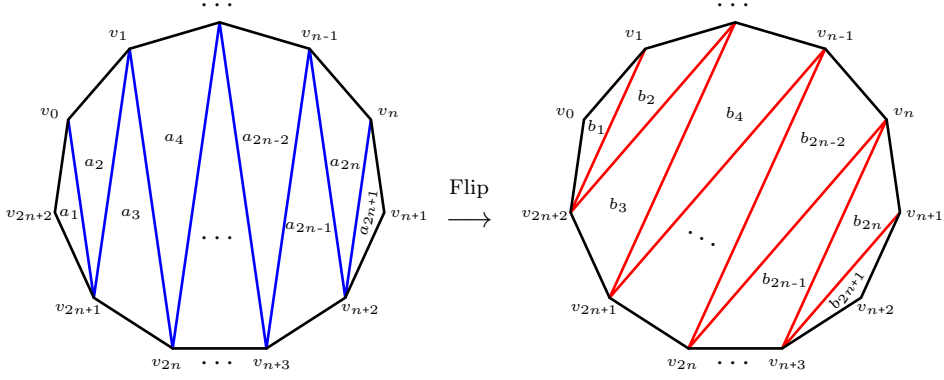
\begin{figure}[tbp]
\centering
\begin{tikzpicture}[scale=2.2, rotate=90]
\usetikzlibrary{calc}

%RHS%
\foreach \i in {1,...,11} {
    \coordinate (P\i) at ({cos(360/11*(\i-1))},{sin(360/11*(\i-1))});
}
\foreach \i in {1,...,11} {
    \coordinate (V\i) at ({1.1*cos(360/11*(\i-1))},{1.12*sin(360/11*(\i-1))});
}

\node at (V1) { $\cdots$};
\node at (V2) {\tiny $v_1$};
\node at (V3) {\tiny $v_0$};
\node at (V4) {\tiny $v_{2n\texttt{+}2}\;\;$};
\node at (V5) {\tiny $v_{2n\texttt{+}1}$};
\node at (V6) {\tiny $v_{2n}$};
\node at (-1.05,0) {$\cdots$};
\node at (V7) {\tiny $v_{n\texttt{+}3}$};
\node at (V8) {\tiny $v_{n\texttt{+}2}$};
\node at (V9) {\tiny $\;\;v_{n\texttt{+}1}$};
\node at (V10) {\tiny $v_{n}$};
\node at (V11) {\tiny $v_{n\texttt{-}1}$};

\draw[line width=1pt,line join=round,red] (P2) -- (P4) -- (P1) -- (P5) -- (P11) -- (P6) -- (P10) -- (P7) -- (P9) ;

\draw[line width=1pt] (P1)
\foreach \i in {2,...,11} { -- (P\i) } -- cycle;

\node[rotate=55] at (-0.6,-0.67) {\scalebox{1}{\tiny $b_{2n\texttt{+}1}$}};
\node at (-0.2,-0.8) {\tiny $b_{2n}$};
\node at (-0.55,-0.3) {\tiny $b_{2n\texttt{-}1}$};
\node at (0.3,-0.53) {\tiny $b_{2n\texttt{-}2}$};
\node at (0.37,0.83) {\tiny $b_{1}$};
\node at (0.55,0.53) {\tiny $b_{2}$};
\node at (-0.1,0.7) {\tiny $b_{3}$};
\node at (0.45,0) {\tiny $b_{4}$};
\node[rotate=-28] at (-0.31,0.19) {$\cdots$};

%LHS%
\begin{scope}[shift={(0,3.1)}]
\foreach \i in {1,...,11} {
    \coordinate (P\i) at ({cos(360/11*(\i-1))},{sin(360/11*(\i-1))});
}
\foreach \i in {1,...,11} {
    \coordinate (V\i) at ({1.1*cos(360/11*(\i-1))},{1.12*sin(360/11*(\i-1))});
}

\draw[line width=1pt,line join=round,blue] (P3) -- (P5) -- (P2) -- (P6) -- (P1) -- (P7) -- (P11) -- (P8) -- (P10);

\node at (-0.3,0) {$\cdots$};

\node at (-0.15,0.9) {\tiny $a_1$};
\node[rotate=80] at (-0.14,-0.9) {\scalebox{1}{\tiny $a_{2n\texttt{+}1}$}};

\node at (-0.15,0.53) {\tiny $a_{3}$};
\node at (-0.23,-0.54) {\tiny $a_{2n\texttt{-}1}$};

\node at (0.15,0.75) {\tiny $a_{2}$};
\node at (0.15,-0.76) {\tiny $a_{2n}$};

\node at (0.3,0.27) {\tiny $a_{4}$};
\node at (0.3,-0.28) {\tiny $a_{2n\texttt{-}2}$};

\node at (V1) { $\cdots$};
\node at (V2) {\tiny $v_1$};
\node at (V3) {\tiny $v_0$};
\node at (V4) {\tiny $v_{2n\texttt{+}2}\;\;$};
\node at (V5) {\tiny $v_{2n\texttt{+}1}$};
\node at (V6) {\tiny $v_{2n}$};
\node at (-1.05,0) {$\cdots$};
\node at (V7) {\tiny $v_{n\texttt{+}3}$};
\node at (V8) {\tiny $v_{n\texttt{+}2}$};
\node at (V9) {\tiny $\;\;v_{n\texttt{+}1}$};
\node at (V10) {\tiny $v_{n}$};
\node at (V11) {\tiny $v_{n\texttt{-}1}$};

\draw[line width=1pt] (P1)
\foreach \i in {2,...,11} { -- (P\i) } -- cycle;

\end{scope}

\node at (0,1.6) {\scriptsize Flip};
\node at (-0.2,1.6) {$\longrightarrow$};

\end{tikzpicture}
\caption{\label{fig: upoly} Two ideal triangulations of $(2n+3)$-gons which are identical up to an unit rotation. A sequence of $2n$ flips of blue edges in the left polygon produces the right one with red edges.
}
\end{figure}
As $k$ varies, the family of modular structures are obtained either from the A-twisted sector of $T[L_{n,k}]$ if it is a rank-0 SCFT, or directly from $T[L_{n,k}]$ if it is a unitary TQFT, and they form a Galois orbit~\cite{Kim:2024dxu,Go:2025ixu,ArabiArdehali:2024ysy,Closset:2026xjj}. 

When $k=n+1$, or $n+2$, the DGG theories $T[L_{n,n+1}]$ and $T[L_{n,n+2}]$ are expected to land on unitary TQFTs whose modular data are related by complex conjugate each other~\cite{Closset:2026xjj}. Indeed, the corresponding two manifolds are related by the orientation reversal
\begin{align}
    L_{n,n+1} = L(2n+3,-1) \setminus \{v\}
    \;\;
    \overset{\text{orientation}}{\longleftrightarrow}
    \;\;
    L_{n,n+2} = L(2n+3,1) \setminus \{v\}
    \,,
\end{align}
thus, we focus only on $T[L_{n,n+1}]$ since $T[L_{n,n+2}]$ is simply the parity conjugate of it.

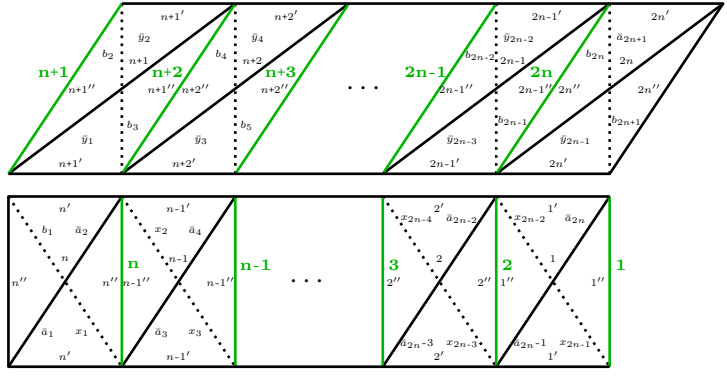
\begin{figure}[tbp]
\centering
\begin{tikzpicture}[scale=1.5, rotate=0]
\usetikzlibrary{calc}

%BOTTOM%
\coordinate (V0) at (0,1.5);
\coordinate (V1) at (1,1.5);
\coordinate (V2) at (2,1.5);
\coordinate (V3) at (3.3,1.5);
\coordinate (V4) at (4.3,1.5);
\coordinate (V5) at (5.3,1.5);
\coordinate (B0) at (0,0);
\coordinate (B1) at (1,0);
\coordinate (B2) at (2,0);
\coordinate (B3) at (3.3,0);
\coordinate (B4) at (4.3,0);
\coordinate (B5) at (5.3,0);

\draw[line width=1pt,line join=round] (V1) -- (B0);
\draw[line width=1pt,line join=round] (V2) -- (B1);
\draw[line width=1pt,line join=round] (V4) -- (B3);
\draw[line width=1pt,line join=round] (V5) -- (B4);

\draw[dotted,line width=1pt,line join=round] (V0) -- (B1);
\draw[dotted,line width=1pt,line join=round] (V1) -- (B2);
\draw[dotted,line width=1pt,line join=round] (V3) -- (B4);
\draw[dotted,line width=1pt,line join=round] (V4) -- (B5);

\draw[green!70!black,line width=1pt,line join=round] (V1) -- (B1);
\draw[green!70!black,line width=1pt,line join=round] (V2) -- (B2);
\draw[green!70!black,line width=1pt,line join=round] (V3) -- (B3);
\draw[green!70!black,line width=1pt,line join=round] (V4) -- (B4);
\draw[green!70!black,line width=1pt,line join=round] (V5) -- (B5);

\draw[line width=1pt,line join=round] (V5) -- (V0) -- (B0) -- (B5) ;

\node at (0.5,1.5-0.1) {\scalebox{0.65}{\tiny $n'$}};
\node at (0.5,0.1) {\scalebox{0.65}{\tiny $n'$}};
\node at (0+0.1,0.75) {\scalebox{0.65}{\tiny $n''$}};
\node at (1-0.1,0.75) {\scalebox{0.65}{\tiny $n''$}};
\node at (0.5,0.95) {\scalebox{0.65}{\tiny $n$}};

\node at (1+0.5,1.5-0.1) {\scalebox{0.65}{\tiny $n\texttt{-}1'$}};
\node at (1+0.5,0.1) {\scalebox{0.65}{\tiny $n\texttt{-}1'$}};
\node at (1+0+0.13,0.75) {\scalebox{0.65}{\tiny $n\texttt{-}1''$}};
\node at (1+1-0.13,0.75) {\scalebox{0.65}{\tiny $n\texttt{-}1''$}};
\node at (1+0.5,0.95) {\scalebox{0.65}{\tiny $n\texttt{-}1$}};

\node at (3.3+0.5,1.5-0.1) {\scalebox{0.65}{\tiny $2'$}};
\node at (3.3+0.5,0.1) {\scalebox{0.65}{\tiny $2'$}};
\node at (3.3+0+0.1,0.75) {\scalebox{0.65}{\tiny $2''$}};
\node at (3.3+1-0.1,0.75) {\scalebox{0.65}{\tiny $2''$}};
\node at (3.3+0.5,0.95) {\scalebox{0.65}{\tiny $2$}};

\node at (4.3+0.5,1.5-0.1) {\scalebox{0.65}{\tiny $1'$}};
\node at (4.3+0.5,0.1) {\scalebox{0.65}{\tiny $1'$}};
\node at (4.3+0+0.1,0.75) {\scalebox{0.65}{\tiny $1''$}};
\node at (4.3+1-0.1,0.75) {\scalebox{0.65}{\tiny $1''$}};
\node at (4.3+0.5,0.95) {\scalebox{0.65}{\tiny $1$}};

%triangles%
\node at (0.35,0.3) {\scalebox{0.65}{\tiny $\bar{a}_1$}};
\node at (0.65,0.3) {\scalebox{0.65}{\tiny $x_1$}};
\node at (0.35,1.2) {\scalebox{0.65}{\tiny $b_1$}};
\node at (0.65,1.2) {\scalebox{0.65}{\tiny $\bar{a}_2$}};

\node at (1+0.35,0.3) {\scalebox{0.65}{\tiny $\bar{a}_3$}};
\node at (1+0.65,0.3) {\scalebox{0.65}{\tiny $x_3$}};
\node at (1+0.35,1.2) {\scalebox{0.65}{\tiny $x_2$}};
\node at (1+0.65,1.2) {\scalebox{0.65}{\tiny $\bar{a}_4$}};

\node at (3.3+0.3,1.3) {\scalebox{0.65}{\tiny $x_{2n\texttt{-}4}$}};
\node at (3.3+0.7,1.3) {\scalebox{0.65}{\tiny $\bar{a}_{2n\texttt{-}2}$}};
\node at (3.3+0.3,0.2) {\scalebox{0.65}{\tiny $\bar{a}_{2n}\texttt{-}3$}};
\node at (3.3+0.7,0.2) {\scalebox{0.65}{\tiny $x_{2n\texttt{-}3}$}};

\node at (4.3+0.3,1.3) {\scalebox{0.65}{\tiny $x_{2n\texttt{-}2}$}};
\node at (4.3+0.7,1.3) {\scalebox{0.65}{\tiny $\bar{a}_{2n}$}};
\node at (4.3+0.3,0.2) {\scalebox{0.65}{\tiny $\bar{a}_{2n}\texttt{-}1$}};
\node at (4.3+0.7,0.2) {\scalebox{0.65}{\tiny $x_{2n\texttt{-}1}$}};

\node at (2.65,0.75) {$\cdots$};

\node at (1+0.1,0.9) {\tiny \textcolor{green!70!black}{${\bf n}$}};
\node at (2+0.18,0.9) {\tiny \textcolor{green!70!black}{$\bf{n\texttt{-}1}$}};
\node at (3.3+0.1,0.9) {\tiny \textcolor{green!70!black}{$\bf{3}$}};
\node at (4.3+0.1,0.9) {\tiny \textcolor{green!70!black}{$\bf{2}$}};
\node at (5.3+0.1,0.9) {\tiny \textcolor{green!70!black}{$\bf{1}$}};

%TOP%
\begin{scope}[shift={(0,1.7)}]

\coordinate (V0) at (1,1.5);
\coordinate (V1) at (2,1.5);
\coordinate (V2) at (3,1.5);
\coordinate (V3) at (4.3,1.5);
\coordinate (V4) at (5.3,1.5);
\coordinate (V5) at (6.3,1.5);
\coordinate (B0) at (0,0);
\coordinate (B1) at (1,0);
\coordinate (B2) at (2,0);
\coordinate (B3) at (3.3,0);
\coordinate (B4) at (4.3,0);
\coordinate (B5) at (5.3,0);

\draw[line width=1pt,line join=round] (V1) -- (B0);
\draw[line width=1pt,line join=round] (V2) -- (B1);
\draw[line width=1pt,line join=round] (V4) -- (B3);
\draw[line width=1pt,line join=round] (V5) -- (B4);

\draw[dotted,line width=1pt,line join=round] (V0) -- (B1);
\draw[dotted,line width=1pt,line join=round] (V1) -- (B2);
\draw[dotted,line width=1pt,line join=round] (V3) -- (B4);
\draw[dotted,line width=1pt,line join=round] (V4) -- (B5);

\draw[green!70!black,line width=1pt,line join=round] (V0) -- (B0);
\draw[green!70!black,line width=1pt,line join=round] (V1) -- (B1);
\draw[green!70!black,line width=1pt,line join=round] (V2) -- (B2);
\draw[green!70!black,line width=1pt,line join=round] (V3) -- (B3);
\draw[green!70!black,line width=1pt,line join=round] (V4) -- (B4);

\draw[line width=1pt,line join=round] (V0) -- (V5) -- (B5) -- (B0) ;

\node at (3.15,0.75) {$\cdots$};

\node at (0.5 + 0.15,0.75) {\scalebox{0.65}{\tiny $n\texttt{+}1''$}};
\node at (1.5 - 0.15,0.75) {\scalebox{0.65}{\tiny $n\texttt{+}1''$}};
\node at (1.15, 1) {\scalebox{0.65}{\tiny $n\texttt{+}1$}};
\node at (0.55, 0.1) {\scalebox{0.65}{\tiny $n\texttt{+}1'$}};
\node at (1.45, 1.4) {\scalebox{0.65}{\tiny $n\texttt{+}1'$}};
\node at (0.4,0.9) {\tiny \textcolor{green!70!black}{${\bf n\texttt{+}1 }$}};
\node at (0.88,1.05) {\scalebox{0.65}{\tiny $b_{2}$}};
\node at (1.2,1.2) {\scalebox{0.65}{\tiny $\bar{y}_{2}$}};
\node at (0.7,0.3) {\scalebox{0.65}{\tiny $\bar{y}_{1}$}};
\node at (1.1,0.43) {\scalebox{0.65}{\tiny $b_{3}$}};

\node at (1+0.5 + 0.15,0.75) {\scalebox{0.65}{\tiny $n\texttt{+}2''$}};
\node at (1+1.5 - 0.15,0.75) {\scalebox{0.65}{\tiny $n\texttt{+}2''$}};
\node at (1+1.15, 1) {\scalebox{0.65}{\tiny $n\texttt{+}2$}};
\node at (1+0.55, 0.1) {\scalebox{0.65}{\tiny $n\texttt{+}2'$}};
\node at (1+1.45, 1.4) {\scalebox{0.65}{\tiny $n\texttt{+}2'$}};
\node at (1+0.4,0.9) {\tiny \textcolor{green!70!black}{${\bf n\texttt{+}2 }$}};
\node at (1+0.88,1.05) {\scalebox{0.65}{\tiny $b_{4}$}};
\node at (1+1.2,1.2) {\scalebox{0.65}{\tiny $\bar{y}_{4}$}};
\node at (1+0.7,0.3) {\scalebox{0.65}{\tiny $\bar{y}_{3}$}};
\node at (1+1.1,0.43) {\scalebox{0.65}{\tiny $b_{5}$}};

\node at (2+0.4,0.9) {\tiny \textcolor{green!70!black}{${\bf n\texttt{+}3 }$}};

\node at (3.3+0.5 + 0.15,0.75) {\scalebox{0.65}{\tiny $2n\texttt{-}1''$}};
\node at (3.3+1.5 - 0.15,0.75) {\scalebox{0.65}{\tiny $2n\texttt{-}1''$}};
\node at (3.3+1.15, 1) {\scalebox{0.65}{\tiny $2n\texttt{-}1$}};
\node at (3.3+0.55, 0.1) {\scalebox{0.65}{\tiny $2n\texttt{-}1'$}};
\node at (3.3+1.45, 1.4) {\scalebox{0.65}{\tiny $2n\texttt{-}1'$}};
\node at (3.3+0.38,0.9) {\tiny \textcolor{green!70!black}{${\bf 2n\texttt{-}1 }$}};
\node at (3.3+0.86,1.04) {\scalebox{0.65}{\tiny $b_{2n\texttt{-}2}$}};
\node at (3.3+1.2,1.2) {\scalebox{0.65}{\tiny $\bar{y}_{2n\texttt{-}2}$}};
\node at (3.3+0.7,0.3) {\scalebox{0.65}{\tiny $\bar{y}_{2n\texttt{-}3}$}};
\node at (3.3+1.15,0.47) {\scalebox{0.65}{\tiny $b_{2n\texttt{-}1}$}};

\node at (4.3+0.5 + 0.15,0.75) {\scalebox{0.65}{\tiny $2n''$}};
\node at (4.3+1.5 - 0.15,0.75) {\scalebox{0.65}{\tiny $2n''$}};
\node at (4.3+1.15, 1) {\scalebox{0.65}{\tiny $2n$}};
\node at (4.3+0.55, 0.1) {\scalebox{0.65}{\tiny $2n'$}};
\node at (4.3+1.45, 1.4) {\scalebox{0.65}{\tiny $2n'$}};
\node at (4.3+0.4,0.9) {\tiny \textcolor{green!70!black}{${\bf 2n }$}};
\node at (4.3+0.88,1.05) {\scalebox{0.65}{\tiny $b_{2n}$}};
\node at (4.3+1.2,1.2) {\scalebox{0.65}{\tiny $\bar{a}_{2n\texttt{+}1}$}};
\node at (4.3+0.7,0.3) {\scalebox{0.65}{\tiny $\bar{y}_{2n\texttt{-}1}$}};
\node at (4.3+1.15,0.47) {\scalebox{0.65}{\tiny $b_{2n\texttt{+}1}$}};
\end{scope}
\end{tikzpicture}
\caption{\label{fig: utri} Minimal triangulation of the punctured lens space $L_{n,n+1}$
}
\end{figure}

\medskip\noindent {\bf Minimal triangulation}.
By the conjecture \eqref{eq: min conj}, the minimal triangulation of $L_{n,n+1}$ will involve $2n$ ideal tetrahedra. Indeed, we find $2n$ flips of internal edges in the triangulation of $(2n+3)$-gon as in Figure \ref{fig: upoly} with the sequence
\begin{align}
    \overline{v_{i} v_{2n+1-i}}
    \;\;\text{for}\; i=0,\cdots,n-1\;,
    \text{then},\;\;
    \overline{v_{i} v_{2n+2-i}}
    \;\;\text{for}\; i=1,\cdots,n
    \,,
\end{align}
where the resulting triangulation is rotated clockwise by one unit.\footnote{This sequence corresponds to the original R-flow of a $\pi$-rotation of the central charge plane for constructing the 4d-3d domain wall theory~\cite{Cecotti:2011iy}, which was recently revisited in the context of the symmetrization map of BPS quivers~\cite{Kucharski:2025lcr}.} The two polygons are identified as $a_i \sim b_i$ with the $2n$ tetrahedra arranged to produce $L(2n+3,1)\setminus\{v\} = L_{n,n+1}$ as desired. Those tetrahedra are shown in Figure \ref{fig: utri} where the lower part arises from the first $n$ flips, while the upper part comes from the remaining $n$ flips. By denoting the faces of the tetrahedra by $a_i$, $b_i$, $x_i$, and $y_i$, with barred notation for the faces on the back side, the $L_{n,n+1}$ is obtained by gluing them as
\begin{align}
    &a_i \sim b_i 
    \;\;
    \text{for}\;
    i=1,\cdots,2n+1\,,
    \nonumber\\
    &x_i \sim y_i
    \;\;
    \text{for}\;
    i=1,\cdots,2n-1
    \,.
\end{align}

\medskip\noindent {\bf The DGG theory}.
By counting the dihedral angle variables meeting at the $i$-th internal edge denoted by green number $i$ in Figure \ref{fig: utri}, the edge variables $C_{i=1,\cdots,2n}$ are written as
\begin{align}
    C_i = 2 Z_{2n+1-i} + (1-\d_{0,[i-1]_n})Z_{i-1} + Z_i
    \,,
\end{align}
from which the coefficient matrices with polarization choice $\s_i=0$ are extracted
\begin{align}
    g_{ij}= 2\d_{i,2n+1-j}
    \;,\;\;
    g'_{ij}= 0
    \;,\;\;
    g''_{ij}= (1-\d_{0,[i-1]_n})\d_{i-1,j} + \d_{i,j}
    \,,
\end{align}
and since $|\det\text{B}|=1$, the DGG theory $T[L_{n,n+1},\s]$ is a 3d $\CN=2$ $U(1)_K^{2n}$ gauge theory having $2n$ diagonally charged chirals $\{\Phi_i\}_{i=1}^{2n}$, with CS level $K$ and superpotential $W$
\begin{align}
    &K_{ij}=\Big\{
    \begin{array}{cc}
        2(-1)^{i+j+1} & \;\;\text{if}\; i+j>2n \;\text{and}\; \text{min}(i,j) \leq n < \text{max}(i,j)  \\
        0 & \;\;\text{otherwise}
    \end{array}
    \nonumber\\
    &W= \sum_{i=1}^{2n} \phi_i^2 V_{\mathfrak{m}^{(i)}}
    \;,\;\;
    \mathfrak{m}_j^{(i)} = -\d_{2n+1-i,j}- ( 1-\d_{0,[-i]_n} ) \d_{2n-i,j}
    \,,\text{for}\; j=1,\cdots,2n\,.
\end{align}
This recovers the theories obtained in \cite{Go:2025ixu} for $n=1,2,3,$ and $4$ from higher power monodromy trace and extends them to arbitrary $n$. We claim that $T[L_{n,n+1},\s]$ flows to a unitary TQFT whose modular data are compatible with the complex conjugate of $(A_1,2n+1)_{\frac{1}{2}}$~\cite{Rowell:2007dge,Schoutens:2015uia,Harvey:2019qzs}.

\subsection{Self-mirror conjecture}
From the orientation reversal $L(p,q) \leftrightarrow L(p,-q)$ combined with the homeomorphism $L(p,q) \sim L(p,q^{-1})$, the lens space is invariant under the orientation reversal if and only if
\begin{align}
    q^2 = -1\,\text{mod}\, p
    \,.
\end{align}
For the punctured lens space $L_{n,k} = L(2n+3,2k)\setminus \{v\}$, this condition reads
\begin{align}
    2k^2 = n+1 \, \text{mod}\, 2n+3
    \,,
    \label{eq: rank-0 cond}
\end{align}
with the first few instances, modulo the orientation reversal, given by
\begin{align}
    (n,k) \in \{ 
    (1,1),(5,4),(7,2),(11,9),(13,6),(17,3),(19,16),(25,15)
    ,\cdots
    \}
    \,,
    \label{eq: cases}
\end{align}
where $(n,k)=(1,1)$ corresponds to the GY theory. Note that the condition \eqref{eq: rank-0 cond} never holds if $k=n+1$ or $k=n+2$, thus, the DGG theory $T[L_{n,k}]$ satisfying \eqref{eq: rank-0 cond} will always be a rank-0 SCFT. The orientation reversal of $L_{n,k}$ is realized as the parity conjugation of $T[L_{n,k}]$ that flips the $U(1)_A$ axial charge, i.e., $\CN=4$ mirror action, thus, we conjecture that the DGG theories associated with orientation-reversal-invariant $L_{n,k}$ are self-mirror:
\begin{align}
    T[L_{n,k}]
    \;\;\text{is a self-mirror rank-0 SCFT if and only if}
    \;\;
    \eqref{eq: rank-0 cond}\;\text{holds.}
    \label{eq: conj}
\end{align}
The superconformal indices~\cite{Kim:2009wb} $\CI_{S^2\times S^1}^{(n,k)}(\eta;q)$ of the first few cases in \eqref{eq: cases} are calculated
\begin{align}
    \CI_{S^2\times S^1}^{(1,1)}(\eta;q) &=
    1 \!-\! q 
    - \Big(\eta+\frac{1}{\eta}\Big) q^{\frac{3}{2}}
    -2 q^2
    - \Big(\eta+\frac{1}{\eta}\Big) q^{\frac{5}{2}}
    -2 q^3 
    - \Big(\eta+\frac{1}{\eta}\Big) q^{\frac{7}{2}}
    -2 q^4 + \cdots
    \nonumber\\
    \CI_{S^2\times S^1}^{(5,4)}(\eta;q) &=
    1 \!-\! q
    \!-\! \Big(\!\eta\!+\!\frac{1}{\eta}\Big) q^{\frac{3}{2}}
    \!-\! q^2
    \!+\! 2\Big(\!\eta\!+\!\frac{1}{\eta}\Big) q^{\frac{5}{2}}
    \!+\! 2\Big(\!3\!+\! \eta^2\!+\!\frac{1}{\eta^2}\Big) q^{3}
    \!+\! 6 \Big(\!\eta \!+\! \frac{1}{\eta}\Big) q^{\frac{7}{2}}
    +\cdots
    \nonumber\\
    \CI_{S^2\times S^1}^{(7,2)}(\eta;q) &=
    1 \!-\! q
    - \Big(\eta+\frac{1}{\eta}\Big) q^{\frac{3}{2}}
    - 2 q^2
    + \Big(1+\eta^2+\frac{1}{\eta^2}\Big) q^{3}
    + 2\Big(\eta+\frac{1}{\eta}\Big) q^{\frac{7}{2}}
    + 3 q^4
    +\cdots
    \nonumber\\
    \CI_{S^2\times S^1}^{(11,9)}(\eta;q) &=
    1 \!-\! q
    \!-\! \Big(\!\eta\!+\!\frac{1}{\eta}\Big) q^{\frac{3}{2}}
    \!-\! q^2
    \!+\! 2\Big(\!\eta\!+\!\frac{1}{\eta}\Big) q^{\frac{5}{2}}
    \!+\! 2\Big(\!3\!+\! \eta^2\!+\!\frac{1}{\eta^2}\Big) q^{3}
    \!+\! 5 \Big(\!\eta \!+\! \frac{1}{\eta}\Big) q^{\frac{7}{2}}
    +\cdots
    \nonumber\\
    \CI_{S^2\times S^1}^{(13,6)}(\eta;q) &=
    1 \!-\! q
    \!-\! \Big(\eta\!+\!\frac{1}{\eta}\Big) q^{\frac{3}{2}}
    \!-\! q^2
    \!+\! 2\Big(\eta\!+\!\frac{1}{\eta}\Big) q^{\frac{5}{2}}
    \!+\! \Big(\!5\!+\! 2\eta^2\!+\!\frac{2}{\eta^2}\Big) q^{3}
    \!+\! \Big(\eta \!+\! \frac{1}{\eta}\Big) q^{\frac{7}{2}}
    +\cdots
    \,,
\end{align}
with convention in~\cite{Gang:2023rei} where $\eta$ is the fugacity of the $U(1)_A$ axial symmetry. We check that
\begin{align}
    \CI_{S^2\times S^1}^{(n,k)}(\eta;q) = \CI_{S^2 \times S^1}^{(n,k)}(\eta^{-1};q)
    \,,
\end{align}
up to $\CO(q^{10})$ which provides strong evidence for the conjecture \eqref{eq: conj}. We provide the 3d $\CN=2$ ACSM descriptions of those DGG theories in appendix \ref{app}. We also conjecture that the associated modular data from the A-/B-twist are related by the complex conjugation.

\section*{Acknowledgements}
We thank Cyril Closset, Hee-Joong Chung, Dongmin Gang, Adam Keyes, Heeyeon Kim, Minsung Kim, and Pietro Longhi for useful discussions. The author acknowledges Cyril Closset for insightful discussions on related projects and for his hospitality during the author's visit to the University of Birmingham. He also acknowledges Pietro Longhi for providing the initial ideas of this project during his visit to Uppsala University. This research is supported by KIAS Individual Grant PG09102 at Korea Institute for Advanced Study.

\appendix
\section{Self-mirror DGG theories} \label{app}
We provide several 3d $\CN=2$ ACSM theories from the DGG construction that flow to the self-mirror rank-0 SCFTs. These theories have as many chiral multiplets as the number of $U(1)$ gauge factors, where the $i$-th one $\Phi_i$ is charged $+\d_{i,j}$ under the $j$-th $U(1)$ and neutral under the R-symmetry. The CS level matrix and the superpotential are denoted by $K$ and $W$ respectively where the latter includes monopole operator terms $\prod_{i} \phi_i^{n_i} V_{\mathfrak{m}}$ with $\phi_i$ being the scalar component of $\Phi_i$ of dressing number $n_i$, while $V_{\mathfrak{m}}$ is the bare monopole operator of magnetic flux $\mathfrak{m}$. By performing the F-maximization~\cite{Jafferis:2010un,Closset:2012vg}, we determine the mixing parameters $\m_i$ of the $U(1)_{T_i}$ topological symmetries and their remaining combination $A$ that becomes the charge of the $U(1)_A$ axial symmetry in the IR.
\begin{itemize}
    \item $T[L_{1,1}]$ : Gang-Yamazaki's minimal rank-0 theory which is a 3d $\CN=2$ $U(1)_K$ gauge theory coupled to a single chiral multiplet $\Phi_1$ with
   \begin{align}
       K = (2)\;,\;\; W=0
       \,.
   \end{align}
   The mixing parameter $\m$ and the charge $A$ are determined as
   \begin{align}
       \m=-1
       \;,\;\;
       A=T_1
       \,.
   \end{align}

   \item $T[L_{5,4}]$ :
    3d $\CN=2$ $U(1)_K^3$ gauge theory coupled to chiral multiplets $\{\Phi_i\}_{i=1}^{3}$ with
   \begin{align}
       K = \left(
       \begin{array}{ccc}
            0 & 2 & 2 \\
            2 & -2 & -2 \\
            2 & -2 & 0
       \end{array}
       \right)
       \;,\;\; 
       W=V_{(2,2,-1)} + \phi_2^2 \phi_3^2 V_{(-1,0,0)}
       \,.
   \end{align}
   The mixing parameter $\m$ and the charge $A$ are determined as
   \begin{align}
       \m=(-2,1,0)
       \;,\;\;
       A = T_2 + 2 T_3
       \,.
   \end{align}

   \item $T[L_{7,2}]$ :
    3d $\CN=2$ $U(1)_K^5$ gauge theory coupled to chiral multiplets $\{\Phi_i\}_{i=1}^{5}$ with
   \begin{align}
       K = \left(
       \begin{array}{ccccc}
            -1 & 2 & 2 & 0 & -2 \\
            2 & 0 & 0 & 0 & 0 \\
            2 & 0 & 2 & 0 & -2 \\
            0 & 0 & 0 & 0 & 2 \\
            -2 & 0 & -2 & 2 & 2
       \end{array}
       \right)
       \;, 
       \begin{array}{c}
        W=\phi_1^2 V_{(0,-1,0,0,0)} + V_{(1,2,-1,0,0)} \\
        \qquad\qquad +\phi_4^2 V_{(0,0,-1,0,-1)} + \phi_5^2 V_{(0,0,0,-1,0)}
       \end{array}
       \,.
   \end{align}
   The mixing parameter $\m$ and the charge $A$ are determined as
   \begin{align}
       \m=(2,-2,-1,-2,-1)
       \;,\;\;
       A = T_1 +  T_3 - T_5
       \,.
   \end{align}

   \item $T[L_{11,9}]$ :
    3d $\CN=2$ $U(1)_K^5$ gauge theory coupled to chiral multiplets $\{\Phi_i\}_{i=1}^{5}$ with
   \begin{align}
       K = \left(
       \begin{array}{ccccc}
            1 & 2 & -2 & 0 & -2 \\
            2 & 0 & 0 & 0 & 0 \\
            -2 & 0 & 2 & 0 & 2 \\
            0 & 0 & 0 & 0 & 2 \\
            -2 & 0 & 2 & 2 & 0
       \end{array}
       \right)
       \;, 
       \begin{array}{c}
        W=\phi_1^2 V_{(0,-1,0,0,0)} + V_{(1,2,2,-1,0)} \\
        \qquad\qquad +\phi_3^2\phi_4^2 V_{(0,-1,0,0,-1)} + \phi_5^2 V_{(0,0,0,-1,0)}
       \end{array}
       \,.
   \end{align}
   The mixing parameter $\m$ and the charge $A$ are determined as
   \begin{align}
       \m=(1,-2,-1,-2,0)
       \;,\;\;
       A = 2 T_1 -  T_3
       \,.
   \end{align}

   \item $T[L_{13,6}]$ :
    3d $\CN=2$ $U(1)_K^5$ gauge theory coupled to chiral multiplets $\{\Phi_i\}_{i=1}^{5}$ with
   \begin{align}
       K = \left(
       \begin{array}{ccccc}
            2 & 0 & -2 & 2 & -2 \\
            0 & 0 & -2 & 2 & -2 \\
            -2 & -2 & -2 & 2 & -2 \\
            2 & 2 & 2 & 0 & 0 \\
            -2 & -2 & -2 & 0 & -1
       \end{array}
       \right)
       \;, 
       \begin{array}{c}
        W=\phi_1^2\phi_2^2\phi_3^2 V_{(0,-1,0,0,0)} + V_{(-1,2,-1,0,0)} \\
        \qquad +\phi_5^2 V_{(0,-1,2,2,0)} + \phi_4^2 V_{(0,0,-1,0,1)}
       \end{array}
       \,.
   \end{align}
   The mixing parameter $\m$ and the charge $A$ are determined as
   \begin{align}
       \m=(-1,0,1,-2,2)
       \;,\;\;
       A = 3 T_1 + 2 T_2 + T_3 + T_5
       \,.
   \end{align}
   
\end{itemize}

\bibliographystyle{JHEP}
\bibliography{ref}

@article{Closset:2026xjj,
    author = "Closset, Cyril and Keyes, Adam and Kim, Sungjoon",
    title = "{Three-dimensional TQFTs from Argyres--Douglas theories via the 3d/3d correspondence}",
    eprint = "2607.20308",
    archivePrefix = "arXiv",
    primaryClass = "hep-th",
    month = "7",
    year = "2026"
}

@article{Jeong:2026dzz,
    author = "Jeong, Kibok and Sohn, Huijoon",
    title = "{Non-unitary Haagerup-like TQFTs and RCFTs from generalized S-fold SCFTs}",
    eprint = "2608.11946",
    archivePrefix = "arXiv",
    primaryClass = "hep-th",
    month = "8",
    year = "2026"
}

@article{Yoshida:2026wvb,
    author = "Yoshida, Yutaka",
    title = "{3d $\mathcal N=4$ rank-zero mirror symmetry, TQFT interfaces, and Zagier duality of Nahm sums}",
    eprint = "2608.08780",
    archivePrefix = "arXiv",
    primaryClass = "hep-th",
    month = "8",
    year = "2026"
}

@article{Chung:2026qyr,
    author = "Chung, Hee-Joong",
    title = "{3d-3d correspondence for knot complements with finite and large $N$}",
    eprint = "2607.21479",
    archivePrefix = "arXiv",
    primaryClass = "hep-th",
    month = "7",
    year = "2026"
}

@article{Chung:2026qfv,
    author = "Chung, Hee-Joong",
    title = "{3d-3d correspondence and abelian flat connection}",
    eprint = "2603.05236",
    archivePrefix = "arXiv",
    primaryClass = "hep-th",
    doi = "10.1007/JHEP07(2026)080",
    journal = "JHEP",
    volume = "07",
    pages = "080",
    year = "2026"
}

@article{Chung:2023qth,
    author = "Chung, Hee-Joong",
    title = "{3d-3d correspondence and 2d $\mathcal{N}$ = (0, 2) boundary conditions}",
    eprint = "2307.10125",
    archivePrefix = "arXiv",
    primaryClass = "hep-th",
    doi = "10.1007/JHEP03(2024)085",
    journal = "JHEP",
    volume = "03",
    pages = "085",
    year = "2024"
}

@article{Chung:2019khu,
    author = "Chung, Hee-Joong",
    title = "{Index for a Model of 3d-3d Correspondence for Plumbed 3-Manifolds}",
    eprint = "1912.13486",
    archivePrefix = "arXiv",
    primaryClass = "hep-th",
    doi = "10.1016/j.nuclphysb.2021.115361",
    journal = "Nucl. Phys. B",
    volume = "965",
    pages = "115361",
    year = "2021"
}

@article{Eckhard:2019jgg,
    author = "Eckhard, Julius and Kim, Heeyeon and Schafer-Nameki, Sakura and Willett, Brian",
    title = "{Higher-Form Symmetries, Bethe Vacua, and the 3d-3d Correspondence}",
    eprint = "1910.14086",
    archivePrefix = "arXiv",
    primaryClass = "hep-th",
    doi = "10.1007/JHEP01(2020)101",
    journal = "JHEP",
    volume = "01",
    pages = "101",
    year = "2020"
}

@article{Gukov:2017kmk,
    author = "Gukov, Sergei and Pei, Du and Putrov, Pavel and Vafa, Cumrun",
    title = "{BPS spectra and 3-manifold invariants}",
    eprint = "1701.06567",
    archivePrefix = "arXiv",
    primaryClass = "hep-th",
    reportNumber = "CALT-TH-2016-039",
    doi = "10.1142/S0218216520400039",
    journal = "J. Knot Theor. Ramifications",
    volume = "29",
    number = "02",
    pages = "2040003",
    year = "2020"
}

@article{Gukov:2016gkn,
    author = "Gukov, Sergei and Putrov, Pavel and Vafa, Cumrun",
    title = "{Fivebranes and 3-manifold homology}",
    eprint = "1602.05302",
    archivePrefix = "arXiv",
    primaryClass = "hep-th",
    reportNumber = "CALT-2016-004",
    doi = "10.1007/JHEP07(2017)071",
    journal = "JHEP",
    volume = "07",
    pages = "071",
    year = "2017"
}

@article{Cheng:2018vpl,
    author = "Cheng, Miranda C. N. and Chun, Sungbong and Ferrari, Francesca and Gukov, Sergei and Harrison, Sarah M.",
    title = "{3d Modularity}",
    eprint = "1809.10148",
    archivePrefix = "arXiv",
    primaryClass = "hep-th",
    reportNumber = "CALT-TH-2018-037",
    doi = "10.1007/JHEP10(2019)010",
    journal = "JHEP",
    volume = "10",
    pages = "010",
    year = "2019"
}

@article{Gang:2018gyt,
    author = "Gang, Dongmin",
    title = "{Quantum Approach to Dehn Surgery Problem}",
    eprint = "1803.11143",
    archivePrefix = "arXiv",
    primaryClass = "math.GT",
    month = "3",
    year = "2018"
}

@article{Chung:2014qpa,
    author = "Chung, Hee-Joong and Dimofte, Tudor and Gukov, Sergei and Sulkowski, Piotr",
    title = "{3d-3d Correspondence Revisited}",
    eprint = "1405.3663",
    archivePrefix = "arXiv",
    primaryClass = "hep-th",
    reportNumber = "CALT-68-2887",
    doi = "10.1007/JHEP04(2016)140",
    journal = "JHEP",
    volume = "04",
    pages = "140",
    year = "2016"
}

@article{Closset:2012vg,
    author = "Closset, Cyril and Dumitrescu, Thomas T. and Festuccia, Guido and Komargodski, Zohar and Seiberg, Nathan",
    title = "{Contact Terms, Unitarity, and F-Maximization in Three-Dimensional Superconformal Theories}",
    eprint = "1205.4142",
    archivePrefix = "arXiv",
    primaryClass = "hep-th",
    reportNumber = "PUPT-2407, WIS-05-12-FEB-DPPA",
    doi = "10.1007/JHEP10(2012)053",
    journal = "JHEP",
    volume = "10",
    pages = "053",
    year = "2012"
}

@article{Harvey:2019qzs,
    author = "Harvey, Jeffrey A. and Hu, Yichen and Wu, Yuxiao",
    title = "{Galois Symmetry Induced by Hecke Relations in Rational Conformal Field Theory and Associated Modular Tensor Categories}",
    eprint = "1912.11955",
    archivePrefix = "arXiv",
    primaryClass = "hep-th",
    doi = "10.1088/1751-8121/ab8e03",
    journal = "J. Phys. A",
    volume = "53",
    number = "33",
    pages = "334003",
    year = "2020"
}

@article{Schoutens:2015uia,
    author = "Schoutens, Kareljan and Wen, Xiao-Gang",
    title = "{Simple-current algebra constructions of 2+1-dimensional topological orders}",
    eprint = "1508.01111",
    archivePrefix = "arXiv",
    primaryClass = "cond-mat.str-el",
    doi = "10.1103/PhysRevB.93.045109",
    journal = "Phys. Rev. B",
    volume = "93",
    number = "4",
    pages = "045109",
    year = "2016"
}

@article{Rowell:2007dge,
    author = "Rowell, Eric and Stong, Richard and Wang, Zhenghan",
    title = "{On Classification of Modular Tensor Categories}",
    eprint = "0712.1377",
    archivePrefix = "arXiv",
    primaryClass = "math.QA",
    doi = "10.1007/s00220-009-0908-z",
    journal = "Commun. Math. Phys.",
    volume = "292",
    number = "2",
    pages = "343--389",
    year = "2009"
}

@article{Kapustin:2010ag,
    author = "Kapustin, Anton and Vyas, Ketan",
    title = "{A-Models in Three and Four Dimensions}",
    eprint = "1002.4241",
    archivePrefix = "arXiv",
    primaryClass = "hep-th",
    month = "2",
    year = "2010"
}

@article{Kim:2025rog,
    author = "Kim, Minsung and Kim, Sungjoon",
    title = "{3D TFTs and boundary VOAs from BPS spectra of (G, G') Argyres-Douglas theories}",
    eprint = "2511.23194",
    archivePrefix = "arXiv",
    primaryClass = "hep-th",
    reportNumber = "KIAS-Q25021",
    doi = "10.1007/JHEP05(2026)203",
    journal = "JHEP",
    volume = "05",
    pages = "203",
    year = "2026"
}

@article{Baek:2024tuo,
    author = "Baek, Seungjoo and Gang, Dongmin",
    title = "{3D bulk field theories for 2D non-unitary $ \mathcal{N} $ = 1 supersymmetric minimal models}",
    eprint = "2405.05746",
    archivePrefix = "arXiv",
    primaryClass = "hep-th",
    doi = "10.1007/JHEP01(2025)027",
    journal = "JHEP",
    volume = "01",
    pages = "027",
    year = "2025"
}

@article{Gang:2026iem,
    author = "Gang, Dongmin and Jeong, Kibok and Kim, Taeyoon and Lee, Soochang",
    title = "{Refined 3D index}",
    eprint = "2604.17449",
    archivePrefix = "arXiv",
    primaryClass = "hep-th",
    month = "4",
    year = "2026"
}

@article{Cho:2020ljj,
    author = "Cho, Gil Young and Gang, Dongmin and Kim, Hee-Cheol",
    title = "{M-theoretic Genesis of Topological Phases}",
    eprint = "2007.01532",
    archivePrefix = "arXiv",
    primaryClass = "hep-th",
    doi = "10.1007/JHEP11(2020)115",
    journal = "JHEP",
    volume = "11",
    pages = "115",
    year = "2020"
}

@article{Baek:2025uev,
    author = "Baek, Seungjoo and Kang, Heesu",
    title = "{Non-hyperbolic 3-manifolds and bulk field theories for supersymmetric/W$_{N}$ minimal models}",
    eprint = "2511.04524",
    archivePrefix = "arXiv",
    primaryClass = "hep-th",
    doi = "10.1007/JHEP03(2026)066",
    journal = "JHEP",
    volume = "03",
    pages = "066",
    year = "2026"
}

@article{Choi:2022dju,
    author = "Choi, Sunjin and Gang, Dongmin and Kim, Hee-Cheol",
    title = "{Infrared phases of 3D class R theories}",
    eprint = "2206.11982",
    archivePrefix = "arXiv",
    primaryClass = "hep-th",
    reportNumber = "KIAS-P22046",
    doi = "10.1007/JHEP11(2022)151",
    journal = "JHEP",
    volume = "11",
    pages = "151",
    year = "2022"
}

@article{Gang:2022kpe,
    author = "Gang, Dongmin and Kim, Dongyeob",
    title = "{Generalized non-unitary Haagerup-Izumi modular data from 3D S-fold SCFTs}",
    eprint = "2211.13561",
    archivePrefix = "arXiv",
    primaryClass = "hep-th",
    doi = "10.1007/JHEP03(2023)185",
    journal = "JHEP",
    volume = "03",
    pages = "185",
    year = "2023"
}

@article{Gang:2024tlp,
    author = "Gang, Dongmin and Kang, Heesu and Kim, Seongmin",
    title = "{Non-hyperbolic 3-manifolds and 3D field theories for 2D Virasoro minimal models}",
    eprint = "2405.16377",
    archivePrefix = "arXiv",
    primaryClass = "hep-th",
    doi = "10.21468/SciPostPhys.20.3.075",
    journal = "SciPost Phys.",
    volume = "20",
    pages = "075",
    year = "2026"
}

@article{Jeong:2025xid,
    author = "Jeong, Kibok and Lee, Soochang",
    title = "{QFT Realization of Non-Unitary $\mathfrak{sl}(2,\mathbb{C})$ WRT Invariants and Their Galois Conjugations}",
    eprint = "2511.16380",
    archivePrefix = "arXiv",
    primaryClass = "hep-th",
    month = "11",
    year = "2025"
}

@article{Garozzo:2019ejm,
    author = "Garozzo, Ivan and Lo Monaco, Gabriele and Mekareeya, Noppadol and Sacchi, Matteo",
    title = "{Supersymmetric Indices of 3d $S$-fold SCFTs}",
    eprint = "1905.07183",
    archivePrefix = "arXiv",
    primaryClass = "hep-th",
    doi = "10.1007/JHEP08(2019)008",
    journal = "JHEP",
    volume = "08",
    pages = "008",
    year = "2019"
}

@article{Assel:2022row,
    author = "Assel, Benjamin and Tachikawa, Yuji and Tomasiello, Alessandro",
    title = "{On $ \mathcal{N} $ = 4 supersymmetry enhancements in three dimensions}",
    eprint = "2209.13984",
    archivePrefix = "arXiv",
    primaryClass = "hep-th",
    doi = "10.1007/JHEP03(2023)170",
    journal = "JHEP",
    volume = "03",
    pages = "170",
    year = "2023"
}

@article{Gang:2018wek,
    author = "Gang, Dongmin and Yonekura, Kazuya",
    title = "{Symmetry enhancement and closing of knots in 3d/3d correspondence}",
    eprint = "1803.04009",
    archivePrefix = "arXiv",
    primaryClass = "hep-th",
    reportNumber = "IPMU-18-0045",
    doi = "10.1007/JHEP07(2018)145",
    journal = "JHEP",
    volume = "07",
    pages = "145",
    year = "2018"
}

@article{Creutzig:2026ajk,
    author = "Creutzig, Thomas and Garner, Niklas and Go, Byeonggi and Kim, Heeyeon",
    title = "{W-algebras of the Deligne-Cvitanovi{\'c} Exceptional series and the minimal 3d ${\mathcal N}=4$ SCFT}",
    eprint = "2603.17394",
    archivePrefix = "arXiv",
    primaryClass = "hep-th",
    month = "3",
    year = "2026"
}

@article{Kim:2025klh,
    author = "Kim, Heeyeon and Kim, Hongseok and Song, Jaewon",
    title = "{Macdonald index from 3d TQFT}",
    eprint = "2511.11186",
    archivePrefix = "arXiv",
    primaryClass = "hep-th",
    doi = "10.1007/JHEP03(2026)213",
    journal = "JHEP",
    volume = "03",
    pages = "213",
    year = "2026"
}

@article{Matveev:1990,
    author        = "Matveev, Sergei V.",
    title         = "{Complexity theory of three-dimensional manifolds}",
    journal       = "Acta Appl. Math.",
    volume        = "19",
    year          = "1990",
    pages         = "101--130",
    doi           = "10.1007/BF00049576"
}

@article{Jaco:2006,
    author        = "Jaco, William and Rubinstein, J. Hyam",
    title         = "{Layered-triangulations of 3-manifolds}",
    eprint        = "math/0603601",
    archivePrefix = "arXiv",
    primaryClass  = "math.GT",
    year          = "2006"
}

@article{Jaco:2009,
    author        = "Jaco, William and Rubinstein, J. Hyam and Tillmann, Stephan",
    title         = "{Minimal triangulations for an infinite family of lens spaces}",
    journal       = "J. Topol.",
    volume        = "2",
    number        = "1",
    year          = "2009",
    pages         = "157--180",
    doi           = "10.1112/jtopol/jtp004",
    eprint        = "0805.2425",
    archivePrefix = "arXiv",
    primaryClass  = "math.GT"
}

@article{Creutzig:2024ljv,
    author = "Creutzig, Thomas and Garner, Niklas and Kim, Heeyeon",
    title = "{Mirror symmetry and level-rank duality for 3d $\mathcal {N} = 4$ rank 0 SCFTs}",
    eprint = "2406.00138",
    archivePrefix = "arXiv",
    primaryClass = "hep-th",
    doi = "10.1007/s11005-025-02015-x",
    journal = "Lett. Math. Phys.",
    volume = "115",
    number = "6",
    pages = "123",
    year = "2025"
}

@article{Ferrari:2023fez,
    author = "Ferrari, Andrea E. V. and Garner, Niklas and Kim, Heeyeon",
    title = "{Boundary vertex algebras for 3d $\mathcal{N}=4$ rank-0 SCFTs}",
    eprint = "2311.05087",
    archivePrefix = "arXiv",
    primaryClass = "hep-th",
    doi = "10.21468/SciPostPhys.17.2.057",
    journal = "SciPost Phys.",
    volume = "17",
    number = "2",
    pages = "057",
    year = "2024"
}

@article{Rozansky:1996bq,
    author = "Rozansky, L. and Witten, Edward",
    title = "{HyperKahler geometry and invariants of three manifolds}",
    eprint = "hep-th/9612216",
    archivePrefix = "arXiv",
    reportNumber = "IASSNS-HEP-96-128",
    doi = "10.1007/s000290050016",
    journal = "Selecta Math.",
    volume = "3",
    pages = "401--458",
    year = "1997"
}

@article{Blau:1996bx,
    author = "Blau, Matthias and Thompson, George",
    title = "{Aspects of N(T) {\ensuremath{>}}= two topological gauge theories and D-branes}",
    eprint = "hep-th/9612143",
    archivePrefix = "arXiv",
    reportNumber = "ENSLAPP-L-630-96, IC-96-262, PAR-LPTHE-96-53",
    doi = "10.1016/S0550-3213(97)00161-2",
    journal = "Nucl. Phys. B",
    volume = "492",
    pages = "545--590",
    year = "1997"
}

@article{Witten:1988ze,
    author = "Witten, Edward",
    title = "{Topological Quantum Field Theory}",
    reportNumber = "IASSNS-HEP-87-72",
    doi = "10.1007/BF01223371",
    journal = "Commun. Math. Phys.",
    volume = "117",
    pages = "353",
    year = "1988"
}

@article{Jafferis:2010un,
    author = "Jafferis, Daniel L.",
    title = "{The Exact Superconformal R-Symmetry Extremizes Z}",
    eprint = "1012.3210",
    archivePrefix = "arXiv",
    primaryClass = "hep-th",
    doi = "10.1007/JHEP05(2012)159",
    journal = "JHEP",
    volume = "05",
    pages = "159",
    year = "2012"
}

@article{Gang:2023rei,
    author = "Gang, Dongmin and Kim, Heeyeon and Stubbs, Spencer",
    title = "{Three-Dimensional Topological Field Theories and Nonunitary Minimal Models}",
    eprint = "2310.09080",
    archivePrefix = "arXiv",
    primaryClass = "hep-th",
    doi = "10.1103/PhysRevLett.132.131601",
    journal = "Phys. Rev. Lett.",
    volume = "132",
    number = "13",
    pages = "131601",
    year = "2024"
}

@article{Nishinaka:2025ytu,
    author = "Nishinaka, Takahiro and Yoshida, Yutaka",
    title = "{3d Chern--Simons matter theories from generalized Argyres--Douglas theories}",
    eprint = "2512.15201",
    archivePrefix = "arXiv",
    primaryClass = "hep-th",
    month = "12",
    year = "2025"
}

@misc{Beem:2014cca,
  author       = {Beem, Christopher},
  title        = {Chiral Symmetry Algebras from Superconformal Symmetry in Four Dimensions},
  howpublished = {Seminar at Crete Center for Theoretical Physics},
  year         = {2014},
  month        = {July},
  note         = {Crete Center for Theoretical Physics}
}

@misc{Beem:2014rka,
  author       = {Beem, Christopher and Rastelli, Leonardo},
  title        = {Infinite Chiral Symmetry in Four and Six Dimensions},
  howpublished = {Seminar at Harvard University},
  year         = {2014},
  month        = {November},
  note         = {Talk by Leonardo Rastelli}
}

@article{Gang:2023ggt,
    author = "Gang, Dongmin and Kim, Dongyeob and Lee, Sungjay",
    title = "{A non-unitary bulk-boundary correspondence: Non-unitary Haagerup RCFTs from S-fold SCFTs}",
    eprint = "2310.14877",
    archivePrefix = "arXiv",
    primaryClass = "hep-th",
    reportNumber = "KIAS-P23052",
    doi = "10.21468/SciPostPhys.17.2.064",
    journal = "SciPost Phys.",
    volume = "17",
    number = "2",
    pages = "064",
    year = "2024"
}

@article{Gang:2021hrd,
    author = "Gang, Dongmin and Kim, Sungjoon and Lee, Kimyeong and Shim, Myungbo and Yamazaki, Masahito",
    title = "{Non-unitary TQFTs from 3D $ \mathcal{N} $ = 4 rank 0 SCFTs}",
    eprint = "2103.09283",
    archivePrefix = "arXiv",
    primaryClass = "hep-th",
    doi = "10.1007/JHEP08(2021)158",
    journal = "JHEP",
    volume = "08",
    pages = "158",
    year = "2021"
}

@article{Gang:2018huc,
    author = "Gang, Dongmin and Yamazaki, Masahito",
    title = "{Three-dimensional gauge theories with supersymmetry enhancement}",
    eprint = "1806.07714",
    archivePrefix = "arXiv",
    primaryClass = "hep-th",
    reportNumber = "IPMU18-0081",
    doi = "10.1103/PhysRevD.98.121701",
    journal = "Phys. Rev. D",
    volume = "98",
    number = "12",
    pages = "121701",
    year = "2018"
}

@article{Dimofte:2012qj,
    author = "Dimofte, Tudor D. and Garoufalidis, Stavros",
    title = "{The Quantum content of the gluing equations}",
    eprint = "1202.6268",
    archivePrefix = "arXiv",
    primaryClass = "math.GT",
    journal = "Geom. Topol.",
    volume = "17",
    pages = "1253--1316",
    year = "2013"
}

@article{NeumannZagier1985,
  author    = {Walter D. Neumann and Don Zagier},
  title     = {Volumes of hyperbolic three-manifolds},
  journal   = {Topology},
  volume    = {24},
  number    = {3},
  pages     = {307--332},
  year      = {1985}
}

@article{Gang:2025ykf,
    author = "Gang, Dongmin and Park, Byoungyoon and Sohn, Huijoon",
    title = "{Torus Knots and Minimal Models Revisited : Rational VOA characters from non-hyperbolic knots}",
    eprint = "2512.23122",
    archivePrefix = "arXiv",
    primaryClass = "hep-th",
    month = "12",
    year = "2025"
}

@article{Dimofte:2011gm,
    author = "Dimofte, Tudor",
    title = "{Quantum Riemann Surfaces in Chern-Simons Theory}",
    eprint = "1102.4847",
    archivePrefix = "arXiv",
    primaryClass = "hep-th",
    doi = "10.4310/ATMP.2013.v17.n3.a1",
    journal = "Adv. Theor. Math. Phys.",
    volume = "17",
    number = "3",
    pages = "479--599",
    year = "2013"
}

@article{Dimofte:2011ju,
    author = "Dimofte, Tudor and Gaiotto, Davide and Gukov, Sergei",
    title = "{Gauge Theories Labelled by Three-Manifolds}",
    eprint = "1108.4389",
    archivePrefix = "arXiv",
    primaryClass = "hep-th",
    reportNumber = "CALT-68-2847",
    doi = "10.1007/s00220-013-1863-2",
    journal = "Commun. Math. Phys.",
    volume = "325",
    pages = "367--419",
    year = "2014"
}

@article{Xie:2012hs,
    author = "Xie, Dan",
    title = "{General Argyres-Douglas Theory}",
    eprint = "1204.2270",
    archivePrefix = "arXiv",
    primaryClass = "hep-th",
    doi = "10.1007/JHEP01(2013)100",
    journal = "JHEP",
    volume = "01",
    pages = "100",
    year = "2013"
}

@article{Kucharski:2025lcr,
    author = "Kucharski, Piotr and Longhi, Pietro and Noshchenko, Dmitry and Park, Sunghyuk and Sulkowski, Piotr",
    title = "{Quivers and BPS states in 3d and 4d}",
    eprint = "2508.09729",
    archivePrefix = "arXiv",
    primaryClass = "hep-th",
    reportNumber = "UUITP-22/25, DIAS-STP-25-19",
    month = "8",
    year = "2025"
}

@article{Closset:2018ghr,
    author = "Closset, Cyril and Kim, Heeyeon and Willett, Brian",
    title = "{Seifert fibering operators in 3d $\mathcal{N}=2$ theories}",
    eprint = "1807.02328",
    archivePrefix = "arXiv",
    primaryClass = "hep-th",
    reportNumber = "CERN-TH-2018-156",
    doi = "10.1007/JHEP11(2018)004",
    journal = "JHEP",
    volume = "11",
    pages = "004",
    year = "2018"
}

@article{Gadde:2013sca,
    author = "Gadde, Abhijit and Gukov, Sergei and Putrov, Pavel",
    editor = "Ballmann, Werner and Blohmann, Christian and Faltings, Gerd and Teichner, Peter and Zagier, Don",
    title = "{Fivebranes and 4-manifolds}",
    eprint = "1306.4320",
    archivePrefix = "arXiv",
    primaryClass = "hep-th",
    reportNumber = "CALT-68-2904",
    doi = "10.1007/978-3-319-43648-7_7",
    journal = "Prog. Math.",
    volume = "319",
    pages = "155--245",
    year = "2016"
}

@article{Dimofte:2011py,
    author = "Dimofte, Tudor and Gaiotto, Davide and Gukov, Sergei",
    title = "{3-Manifolds and 3d Indices}",
    eprint = "1112.5179",
    archivePrefix = "arXiv",
    primaryClass = "hep-th",
    doi = "10.4310/ATMP.2013.v17.n5.a3",
    journal = "Adv. Theor. Math. Phys.",
    volume = "17",
    number = "5",
    pages = "975--1076",
    year = "2013"
}

@article{Kim:2009wb,
    author = "Kim, Seok",
    title = "{The Complete superconformal index for N=6 Chern-Simons theory}",
    eprint = "0903.4172",
    archivePrefix = "arXiv",
    primaryClass = "hep-th",
    reportNumber = "IMPERIAL-TP-09-SK-01",
    doi = "10.1016/j.nuclphysb.2009.06.025",
    journal = "Nucl. Phys. B",
    volume = "821",
    pages = "241--284",
    year = "2009",
    note = "[Erratum: Nucl.Phys.B 864, 884 (2012)]"
}

@article{Cordova:2015nma,
    author = "Cordova, Clay and Shao, Shu-Heng",
    title = "{Schur Indices, BPS Particles, and Argyres-Douglas Theories}",
    eprint = "1506.00265",
    archivePrefix = "arXiv",
    primaryClass = "hep-th",
    doi = "10.1007/JHEP01(2016)040",
    journal = "JHEP",
    volume = "01",
    pages = "040",
    year = "2016"
}

@article{Beem:2017ooy,
    author = "Beem, Christopher and Rastelli, Leonardo",
    title = "{Vertex operator algebras, Higgs branches, and modular differential equations}",
    eprint = "1707.07679",
    archivePrefix = "arXiv",
    primaryClass = "hep-th",
    reportNumber = "YITP-SB-17-27",
    doi = "10.1007/JHEP08(2018)114",
    journal = "JHEP",
    volume = "08",
    pages = "114",
    year = "2018"
}

@article{Kim:2024dxu,
    author = "Kim, Heeyeon and Song, Jaewon",
    title = "{A family of vertex algebras from Argyres-Douglas theory}",
    eprint = "2412.20015",
    archivePrefix = "arXiv",
    primaryClass = "hep-th",
    doi = "10.21468/SciPostPhys.19.6.144",
    journal = "SciPost Phys.",
    volume = "19",
    number = "6",
    pages = "144",
    year = "2025"
}

@article{Gaiotto:2009hg,
    author = "Gaiotto, Davide and Moore, Gregory W. and Neitzke, Andrew",
    title = "{Wall-crossing, Hitchin systems, and the WKB approximation}",
    eprint = "0907.3987",
    archivePrefix = "arXiv",
    primaryClass = "hep-th",
    doi = "10.1016/j.aim.2012.09.027",
    journal = "Adv. Math.",
    volume = "234",
    pages = "239--403",
    year = "2013"
}

@article{Cecotti:2011iy,
    author = "Cecotti, Sergio and Cordova, Clay and Vafa, Cumrun",
    title = "{Braids, Walls, and Mirrors}",
    eprint = "1110.2115",
    archivePrefix = "arXiv",
    primaryClass = "hep-th",
    month = "10",
    year = "2011"
}

@article{Gang:2024loa,
    author = "Gang, Dongmin and Kim, Heeyeon and Park, Byoungyoon and Stubbs, Spencer",
    title = "{Three dimensional topological field theories and Nahm sum formulas}",
    eprint = "2411.06081",
    archivePrefix = "arXiv",
    primaryClass = "hep-th",
    doi = "10.21468/SciPostPhys.19.5.128",
    journal = "SciPost Phys.",
    volume = "19",
    number = "5",
    pages = "128",
    year = "2025"
}

@article{Go:2025ixu,
    author = "Go, Byeonggi and Jia, Qiang and Kim, Heeyeon and Kim, Sungjoon",
    title = "{From BPS spectra of Argyres-Douglas theories to families of 3d TFTs}",
    eprint = "2502.15133",
    archivePrefix = "arXiv",
    primaryClass = "hep-th",
    doi = "10.1007/JHEP08(2025)012",
    journal = "JHEP",
    volume = "08",
    pages = "012",
    year = "2025"
}

@article{Gaiotto:2024ioj,
    author = "Gaiotto, Davide and Kim, Heeyeon",
    title = "{3D TFTs from 4d $ \mathcal{N} $ = 2 BPS particles}",
    eprint = "2409.20393",
    archivePrefix = "arXiv",
    primaryClass = "hep-th",
    doi = "10.1007/JHEP03(2025)173",
    journal = "JHEP",
    volume = "03",
    pages = "173",
    year = "2025"
}

@article{ArabiArdehali:2024vli,
    author = "Arabi Ardehali, Arash and Gang, Dongmin and Rajappa, Neville Joshua and Sacchi, Matteo",
    title = "{3d SUSY enhancement and non-semisimple TQFTs from four dimensions}",
    eprint = "2411.00766",
    archivePrefix = "arXiv",
    primaryClass = "hep-th",
    reportNumber = "YITP-SB-2024-26",
    doi = "10.1007/JHEP09(2025)179",
    journal = "JHEP",
    volume = "09",
    pages = "179",
    year = "2025"
}

@article{ArabiArdehali:2024ysy,
    author = "Arabi Ardehali, Arash and Dedushenko, Mykola and Gang, Dongmin and Litvinov, Mikhail",
    title = "{Bridging 4D QFTs and 2D VOAs via 3D high-temperature EFTs}",
    eprint = "2409.18130",
    archivePrefix = "arXiv",
    primaryClass = "hep-th",
    reportNumber = "YITP-SB-2024-13",
    month = "9",
    year = "2024"
}

@article{Beem:2013sza,
    author = "Beem, Christopher and Lemos, Madalena and Liendo, Pedro and Peelaers, Wolfger and Rastelli, Leonardo and van Rees, Balt C.",
    title = "{Infinite Chiral Symmetry in Four Dimensions}",
    eprint = "1312.5344",
    archivePrefix = "arXiv",
    primaryClass = "hep-th",
    reportNumber = "YITP-SB-13-45, CERN-PH-TH-2013-311, HU-EP-13-78",
    doi = "10.1007/s00220-014-2272-x",
    journal = "Commun. Math. Phys.",
    volume = "336",
    number = "3",
    pages = "1359--1433",
    year = "2015"
}

@article{Dedushenko:2023cvd,
    author = "Dedushenko, Mykola",
    title = "{On the 4d/3d/2d view of the SCFT/VOA correspondence}",
    eprint = "2312.17747",
    archivePrefix = "arXiv",
    primaryClass = "hep-th",
    month = "12",
    year = "2023"
}

@article{Closset:2019hyt,
    author = "Closset, Cyril and Kim, Heeyeon",
    title = "{Three-dimensional $\mathcal{N}$ = 2 supersymmetric gauge theories and partition functions on Seifert manifolds: A review}",
    eprint = "1908.08875",
    archivePrefix = "arXiv",
    primaryClass = "hep-th",
    doi = "10.1142/S0217751X19300114",
    journal = "Int. J. Mod. Phys. A",
    volume = "34",
    number = "23",
    pages = "1930011",
    year = "2019"
}
\end{document}